\documentclass[lettersize,journal]{IEEEtran}

\usepackage{url}
\usepackage{verbatim}

\usepackage{cite}
\usepackage{amsmath,amssymb,amsfonts}
\usepackage{mathrsfs}
\usepackage[linesnumbered, ruled]{algorithm2e}
\usepackage{amsthm}
\usepackage{multirow}
\usepackage{multicol}
\usepackage{tabularx}
\usepackage{algorithmic}
\usepackage{booktabs}
\usepackage{graphicx}
\usepackage{float}  
\usepackage{subfigure}  
\usepackage{bm}
\usepackage{booktabs}
\usepackage{textcomp}
\usepackage{xcolor}
\begin{document}
	
	\title{Cross-Layer Optimization for Statistical QoS Provision in C-RAN with Finite-Length Coding\\}
	
	\author{Chang Wu, Hancheng Lu,~\IEEEmembership{Senior Member,~IEEE}, Yuang Chen,~\IEEEmembership{Student Member,~IEEE}, and Langtian Qin
		\thanks{Chang Wu, Hancheng Lu, Yuang Chen and Langtian Qin are with  CAS Key Laboratory of Wireless-Optical Communications, University of Science and Technology of China, Hefei 230027, China (e-mail: changwu@mail.ustc.edu.cn; hclu@ustc.edu.cn; \{yuangchen21, qlt315\}@mail.ustc.edu.cn). Hancheng Lu is also with Institute of Artificial Intelligence, Hefei Comprehensive National Science Center, Hefei 230088, China.}
		
	}

	
	
	\maketitle
	\begin{abstract}
		The cloud radio access network (C-RAN) has become the foundational structure for various emerging communication paradigms, leveraging the flexible deployment of distributed access points (APs) and centralized task processing. In this paper, we propose a cross-layer optimization framework based on a practical finite-length coding communication system in C-RAN, aiming at maximizing bandwidth efficiency while providing statistical quality of service (QoS) for individual services. Based on the theoretical results from effective capacity and finite-length coding, we formulate a joint optimization problem involving modulation and coding schemes (MCS), retransmission count, initial bandwidth allocation and AP selection, which reflects the coordinated decision of parameters across the physical layer, data link layer and transport layer. To tackle such a mixed-integer nonlinear programming (MINLP) problem, we firstly decompose it into a transmission parameter decision (TPD) sub-problem and a user association (UA) sub-problem, which can be solved by a binary search-based algorithm and an auction-based algorithm respectively. Simulation results demonstrate that the proposed model can accurately capture the impact of QoS requirements and channel quality on the optimal transmission parameters. Furthermore, compared with fixed transmission parameter setting, the proposed algorithms achieve the bandwidth efficiency gain up to 27.87\% under various traffic and channel scenarios.
	\end{abstract}
	
	
	\newtheorem{def1}{\bf Definition}
	\newtheorem{thm1}{\bf Theorem}
	\newtheorem{lem1}{\bf Lemma}
	\newtheorem{cor1}{\bf Corollary}

	\section{Introduction}
	\IEEEPARstart{T}{he} advent of the fifth-generation and beyond (5G/B5G) mobile communication technology has paved the way for emerging communication paradigms and innovative services, such as industrial automation, virtual reality (VR), remote training \cite{morgado2018survey}, \cite{navarro2020survey}. The rapid expansion of these services is in turn providing a fertile ground for further advancements in radio technology, driven by the escalating demand for enhanced connectivity. As the crucial ``last mile'' of data delivery, the radio access network (RAN) assumes critical significance in meeting the stringent requirements of these services \cite{haile2021end}, particularly when operating alongside ultra-high-speed wired links, thus circumventing the limitations imposed by user cables. Consequently, novel radio technologies and network paradigms have been developed, such as massive multiple-input-multiple-output (massive MIMO) and user-centric networks (UCN) \cite{ammar2021user}. Among them, the 5G cloud RAN (C-RAN) has garnered considerable attention from both academia and industry, owing to its distinctive deployment structure and significant commercial potential \cite{hossain2019recent,suryaprakash2015heterogeneous}. On the one hand, it leverages cloud computing and virtualization technologies to decouple the base station (BS) into the baseband unit (BBU) and the remote radio head (RRH). The BBU is responsible for baseband signal processing, while the RRH focuses on signal amplification and modulation. This centralization of computing units, combined with the distributed deployment of radio frequency (RF) units, forms the technological underpinning for various emerging technologies, such as mobile edge computing (MEC) and coordinated multipoint (CoMP) transmission \cite{qin2022user,moosavi2021delay,elhattab2020comp}. On the other hand, the sharing of BBU pools among multiple operators through computing unit rentals presents an effective approach to reduce operational costs \cite{suryaprakash2015heterogeneous}.
	
	Considering the immense potential and the derivative architectures of C-RAN, extensive research has been conducted in recent years to propose enhanced control schemes at each layer of the system. The throughput is maximized through joint optimization in \cite{al2020throughput,lee2015message,Al2019crosslayercloud,Al2022ajointrein}, considering constraints such as RRH associations, transmit power and bandwidth allocation. In the context of green communications, the minimization of transmit power or maximization of energy efficiency under specific service requirements has been extensively studied in \cite{iqbal2021double,tang2015cross,she2017cross,li2016energy,tham2017energy,he2014energy,pradhan2020computation}. Authors in \cite{elhattab2020comp,tang2019systematic,liu2019economically} integrate various system benefits into utility functions to achieve joint optimization of system performance. Nevertheless, many works primarily focus on pursuing optimal performance based on given system resources without considering the specific QoS requirements of individual services \cite{elhattab2020comp,tang2015cross,al2020throughput,lee2015message,she2017cross,ferdouse2019joint,pradhan2020computation}. Moreover, other optimization efforts that take into account traffic characteristics have not fully characterized the utilization of underlying resources \cite{liu2019economically,tang2019systematic}. In addition, the majority of studies make decisions on user scheduling and power, bandwidth, and computational resource allocation based on ideal infinite-length channel codes, where Shannon capacity is considered as the actual throughput of users \cite{iqbal2021double,elhattab2020comp,tang2015cross,al2020throughput,ferdouse2019joint,lee2015message,she2017cross,li2016energy,tang2019systematic,Al2019crosslayercloud,Al2022ajointrein,pradhan2020computation}. Only a few references consider practical finite-length coding communication schemes and the associated bit error rate \cite{liu2019economically,tham2017energy,he2014energy}, but in-depth analyses are lacking \cite{tham2017energy,he2014energy}. Remarkably, performance analyses based on Shannon capacity not only overestimate the actual performance of communication systems but also lack analysis of the impact of transmission parameters on service performance in finite-length channel coding communication, which is the obstacle we aim to overcome in this work.
	
	Aforementioned researches only optimize the transmission in one or two layers of the system, which has natural limitations compared with cross-layer optimization that jointly considers the upper-layer applications and the lower-layer resources. In other words, applications with diverse quality-of-service (QoS) requirements drive optimization across multiple layers. The ultra-reliable low-latency communication (uRLLC) services, for example, require strong reliability ($\sim$99.999\%) and ultra-low latency ($\sim$1ms), while the enhanced mobile broadband (eMBB) applications expect high capacity with some tolerance for loss \cite{navarro2020survey}. If the same robustness guarantee and latency tolerance is applied uniformly to all types of business, some will fail to fulfill the requirements, while others will waste scarce wireless resources. On the other hand, the distributed RF units provide a structural foundation for implementing coordinated and collaborative transmission/reception strategies between RRHs, thereby improving spectrum efficiency and energy efficiency \cite{hossain2019recent,iqbal2021double,elhattab2020comp,tang2015cross}. However, just as there is no free lunch, efficient User Association (UA) and resource allocation algorithms have to be carefully studied to fully realize the potential of system.

	
	To overcome the aforementioned challenges, the relationship between the QoS requirements of the applications and the transmission parameters in the lower layer is analyzed and correlated from cross-layer perspective. In view of the rigid transmission parameter decision (TPD) in the current communication system, i.e., the fixed bit error rate (BER) threshold used to derive MCS \cite{3gpp_technical_38214} and the maximum retransmission count in automatic retransmission request (ARQ) protocol \cite{ahmed2021hybrid}, we design a flexible TPD algorithm to achieve the minimum bandwidth consumption under individual services with statistical QoS requirements, i.e., the maximum bandwidth efficiency. Deterministic QoS provision is typically
	hard due to the high volatility of wireless channels \cite{wu2003effective,zhang2018heterogeneous,chen2023streaming,chen2023statistical,Yuang2023When}. Therefore, the statistical delay QoS provision is characterized based on effective capacity theory \cite{musavian2015effective,guo2019resource}. The main contributions of our work can be summarized as follows:

	\begin{itemize}
		\item We propose a QoS provision framework for finite-length coding communication systems in C-RAN. This framework jointly controls the transmission parameters at each layer to provide statistical QoS guarantee for diverse services, thereby leading to maximum bandwidth efficiency. The framework takes into account queuing delay and transmission delay constraints, as well as data loss rate constraints caused by queuing timeouts and transmission errors, forming an optimization space that includes bandwidth allocation for initial transmission, UA in the transport layer, maximum transmission count in the link layer, and MCS in the physical layer. The data loss rate due to queuing timeout is represented by the latency violation probability (LVP) in the effective capacity theory.
		\item Considering the complexity of the problem, we decompose the original optimization problem into TPD sub-problem within each AP-user pair and UA decision sub-problem, which can be solved respectively by a binary search-based algorithm and an auction-based algorithm. The algorithm can ultimately find the global optimal solution of the original problem with polynomial time complexity.
		\item Extensive numerical simulations are conducted to validate the rationality of the finite-length coding system and the effective capacity model, as well as the effectiveness of our proposed algorithms in terms of statistical QoS provision and bandwidth efficiency improvement. The algorithms achieve maximum bandwidth efficiency while satisfying the requirements of statistical QoS provision.
	\end{itemize}
	
	The rest of this paper is organized as follows. In Section \ref{SysM}, we formulate a bandwidth consumption minimization problem in C-RAN based on the effective capacity theory and the finite-length coding model. In Section \ref{ProbSolu}, parameter decoupling is implemented and algorithms are designed to address two sub-problems separately, aiming to solve the initial optimization problem. Simulation results of the proposed algorithm are discussed in Section \ref{Performance E}. Finally, Section \ref{Conclusion} concludes the whole paper and gives future research directions.

	\section{System Model and Problem Formulation}\label{SysM}
	\begin{figure}[tb]
		\centering
		\includegraphics[scale=0.4]{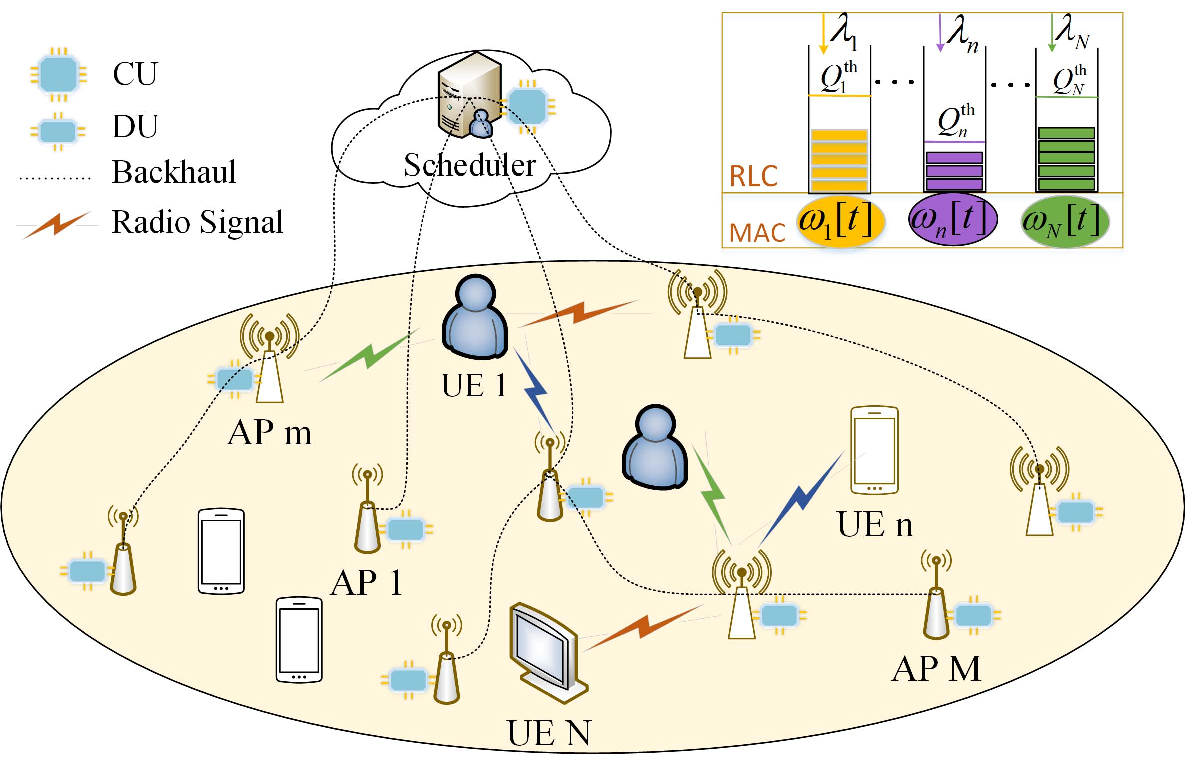}
		\caption{C-RAN system with one buffer queue per user flow.}
		\label{Fig:Scenario}
	\end{figure}
	
	This section will first present the network model, service model and the AMC model, and then formulate a joint optimization problem aiming to maximize the bandwidth efficiency of the C-RAN system subject to constraints on the delay and loss QoS requirements.
	
	\subsection{Network Model}
	As shown in Fig. \ref{Fig:Scenario}, $N$ users are collaboratively served by $M$ radio access points (RAPs) in the C-RAN system, where the set of user indexes and the AP indexes are denoted as $\mathcal{N}=\{1,2,\cdots,N\}$ and $\mathcal{M}=\{1,2,\cdots,M\}$, respectively. All APs are controlled by a scheduler located in the Central Unit (CU). We consider each user requesting a downlink real-time data flow with certain QoS requirements, such as multimedia teleconferencing and live. Considering the fluctuating channel capacity arising from the dynamic nature of wireless channels, the buffer is configured to optimize the utilization of channel resources. The data packet from the source server first enters the RLC buffer allocated for each user in the CU to wait for service. Assuming that the data of user $n \in \mathcal{N}$ enters the corresponding queue at a constant rate $\lambda_n$.
	
	The media access control (MAC) scheduler in CU allocates a certain amount of transmission opportunities to the data queues in the RLC queue based on the connection between APs and users during each scheduling cycle. As a result, data units are read from the queue header and assembled into a transmission block (TB) for transmission, and ARQ are applied to improve the robustness of the delivery. The radio air-port delay of the data consists of the queuing delay in the RLC queue, the link transmission delay from CU to APs, and the service delay at the AP. Among them, the queuing delay caused by the mismatch between the arrival rate and the service rate, as well as the the service delay (including transmission and retransmission delay) dominates the air-port delay, which are performance requirements we need to meet in this work. In order to fulfill the low-latency requirement of real-time data flows and the distortion limitation caused by data loss, it is necessary to study the collaborative optimization of UA and the parameter configuration of APs.
	
	\subsection{Service Model}
	The wireless channels between users and APs are characterized as block fading, implying that the channel response remains constant in the duration of each fading block, assuming a slot with length $T^{\text{s}}$, and varies randomly across different fading blocks. With a constant arrival rate $\lambda_n$, we define the data arrival process of user $n$ during $[0,t)$ as $A_n(t) = \lambda_n \cdot t$. Let $[t]$ denote the $t$-th slot and $(t)$ denote the moment $t$. Similarly, the cumulative service process provided by the C-RAN system for user $n$ is represented as
	\begin{equation}
		\widetilde{S}_{n}(t) = S_{n}\left( t-t\%T^{s} \right) + \left( t\%T^{s} \right) \omega_{n}\left[t/T^{s} + 1\right],
	\end{equation}
	where operator $\%$ denotes modulo operation, and $\omega_{n}[t]$ is the average service rate that the system could provide for user $n$ in the $t$-th slot, which may be greater than the rate required. So the actual service process accepted by user $n$ is represented as $S_{n}(t) = \min \left\{\widetilde{S}_{n}(t), A_{n}(t) \right\}$. With the average service rate $\omega_{m,n}[t]$ provided by AP $m \in \mathcal{M}$ for user $n$ at slot $t$, the total service rate can be represented as
	\begin{equation}
		\omega_{n}\left[ t \right] = \sum_{m \in \mathcal{M}} a_{m,n}\omega_{m,n} [t],
	\end{equation}
	where $a_{m,n}$ is UA indicator regarding the AP $m$ and  the user $n$. In particular, $a_{m,n}=1$ when the $n$th user is associated with the $m$th AP, and $a_{m,n}=0$ otherwise.
	
	Due to the diversity of QoS requirements for different traffic flows, such as latency and data loss rate, as well as the varying degree of importance for the same content by different users, we can flexibly adjust the service process provided for each service flow according to its characteristics. By doing so, we can maximize the bandwidth efficiency of the entire system within its acceptable QoS loss constraint. We assume that the maximum tolerable queuing delay for user $n$ is $D_{n}^{\text{q,th}}$. The scheduler will discard the part of the queue that exceeds the delay threshold, with a maximum allowable queue length $Q_{n}^{\text{th}} = \lambda_{n} D_{n}^{\text{q,th}}$. Research shows that the LVP of user $n$ can be represented as the ratio of the amount of data served in the violating time domain to the total service data when the slot length is sufficiently large \cite{guo2019resource,musavian2015effective}.

	Based on the above analysis, the service process of the system across different slot is uncorrelated. Therefore, the effective capacity of the service system for user $n$ can be represented as \cite{musavian2015effective}
	\begin{equation}\label{EC0}
		EC_{n}\left( \theta_{n} \right) = -\frac{1}{\theta_{n}T^{\text{s}}} \ln{\mathbb{E}_{\gamma}\left\{ e^{-\theta_{n}T^{\text{s}}\omega_{n}[t]} \right\}},
	\end{equation}
	where $\mathbb{E}_{\gamma}{\left\{ \cdot \right\}}$ indicates the statistical average of inner arguments with respect to service rate, and $\theta_{n} \geq 0$ is the latency exponent of the $n$th user, which indicates the tolerance of a satisfactory system to the occurrence of delay violations. According to \cite{guo2019resource}, $EC_{n}\left( \theta_{n} \right) $ is a monotonic descending function with respect to $\theta_{n}$ when the service process of the system $\omega_{n}[t]$ is determined. Therefore, the system does not guarantee the LVP when the effective capacity gets the maximum value $EC_{n}\left( \theta_{n} = 0 \right) = \mathbb{E}_{\gamma}\left\{ \omega_{n}[t] \right\}$, and $EC_{n}\left( \theta_{n} = \infty \right) = 0$ indicates that the system cannot tolerate any possible delay violations, resulting in zero capacity. The gap between $EC_{n}\left( \theta_{n} \right) $ and the provided capacity provides a delay violation guarantee with parameter $\theta_{n}$. In other words, the effective capacity provides the maximum supported source rate for a given service rate under the delay violation probability guarantee characterized by parameter $\theta_{n}$. When the source rate satisfies $\lambda_{n} = EC_{n}\left( \theta_{n} \right)$, the probability of violation of the delay threshold can be approximated as
	\begin{equation} \label{LVP0}
		\Pr\left\{ D_{n}^{\text{q}} > D_{n}^{\text{q,th}} \right\} \approx \varphi_{n} \cdot e^{-\theta_{n} \lambda_{n} D_{n}^{\text{q,th}}},
	\end{equation}
	where $\varphi_{n}$ is the non-empty probability of the queue that can be approximated as $\varphi_{n} \approx \frac{\lambda_{n}}{\mathbb{E}_{\gamma}\left\{\omega_{n}[t] \right\}} < 1$, and $D_{n}^{q}$ is a random variable representing the steady-state queuing delay experienced by the data of user $n$ before it is served.
	
	\subsection{Modulation and Channel Coding Model}
	Considering the gap between the theoretical Shannon rate and the actual performance \cite{qiu1999performance,seo2004proportional}, with the coefficient $\varrho =-2\ln\left( 5\cdot \rho \right) / 3$, the relationship between the maximum available spectral efficiency and the signal-to-interference-noise-ratio (SINR) of the channel can be expressed as $\hat{\upsilon} = \log_{2}\left( 1+ \gamma / \varrho \right)$, where SINR $\gamma$ is the unique parameter used to describe channel quality in this article. Therefore, we select the highest MCS index under the allowed BER $\rho$, with the MCS switching threshold being represented as \cite{seo2004proportional}
	\begin{equation}\label{switch threshold SNR}
		\gamma_{j} = \frac{2}{3}\left(1-2^{\upsilon_{j}}\right)\ln\left(5\cdot\rho\right), j \in \left\{0,1,\cdots,J\right\},
	\end{equation}
	where $\upsilon_{j}$ is the spectral efficiency when MCS $j$ is selected. And $J$ is the maximum MCS index that can be selected, so we have $\gamma_{_{J+1}}=+\infty$. Meanwhile, the actual BER can be calculated through channel SINR and selected MCS, which can be represented as
	\begin{equation} \label{actual BER}
		Pb(j, \gamma)=\frac{1}{5} \exp \left(-\frac{1.5 \cdot \gamma}{2^{\upsilon_{j}}-1}\right).
	\end{equation}
	We adopt Rayleigh channel model to describe $\gamma$ statistically. The SINR $\gamma$ for each fading block is thus a random variable with a probability density function (pdf):
	\begin{equation}\label{SINR pdf}
		p_{\gamma}(\gamma) = \frac{1}{\bar{\gamma}}e^{-\frac{\gamma}{\bar{\gamma}}},
	\end{equation}
	where $\bar{\gamma}=\mathbb{E}_{\gamma}\left[\gamma\right]$ is the average SINR related to the distance between APs and users.
	
	In NR systems, data stored in RLC service data unit (SDU) format is encapsulated, segmented, and assembled into TBs when the MAC scheduler notifies the queue of a transmission opportunity. Then, the oversized TB will be divided into smaller code blocks in the channel coding process. Code blocks have the same size due to the appropriate TB size design \cite{3gpp_technical_38214}, which are the basic unit of channel coding and rate matching. With the maximum code block size $L$ (bit) and AMC-induced BER $P_b$, per-transmission block error rate (BLER) can be calculated as $P(j, \gamma)=1-\left[1-Pb(j, \gamma)\right]^{L}$. For the ergodic channel, the average BLER at the physical layer can be calculated as 
	\begin{equation}\label{average Pcb}
		\begin{aligned}
			\bar{P}
			&=\frac{1}{P_{T}} \sum_{j=0}^{J} \int_{\gamma_{j}}^{\gamma_{j+1}} P\left(j, \gamma\right) p_{\gamma}\left(\gamma\right) d\gamma \\
			&=\frac{1}{P_{T}} \sum_{j=0}^{J} \int_{\gamma_{j}}^{\gamma_{j+1}} \left[ 1\!\!-\!\!\left[1\!\!-\!\!Pb(j, \gamma)\right]^{L} \right] p_{\gamma}\left(\gamma\right) d\gamma,
		\end{aligned}
	\end{equation}
	where $P_{T} = \int_{\gamma_{0}}^{+\infty} p_{\gamma}\left(\gamma\right) d\gamma$ is the probability that the channel has no deep fading and at least one MCS mode is available \cite{wu2007cross}. 
	By utilizing $X$ transmissions of code blocks by ARQ, the actual BLER at the link layer can be calculated as $P^{\text{tr}} = \bar{P}^{X}$. Furthermore, we assume that the block can be successfully decoded when the actual BLER falls below threshold $\varepsilon_{0}$ due to the cooperation with forward error correction (FEC) coding \cite{hagenauer1987forward,mizuochi2006recent,paul2015implementation}.
	
	\subsection{Problem Formulation}\label{Sec: ProbFo}
	For simplicity, orthogonal frequency division multiplexing (OFDM) is adopted between APs and users to avoid interference, which indicates that the SINR in the previous text can be replaced by signal-to-noise-ratio (SNR). The scheduling slot period $T^{\text{s}}$ is obtained while selecting the subcarrier spacing according to the numerology. According to the NR system, the resource block (RB) is the basic unit for wireless resource allocation, and any number of RBs not exceeding the total resources of system can be allocated for each transmission. We assume that each RB contains $\alpha$ subcarrier and $\beta$ OFDM symbols with spectral efficiency $\upsilon_{j}$ in the C-RAN system, then the amount of information that can be transmitted by an RB in MCS $j$ mode can be expressed as $\psi_{j} = \alpha \beta \upsilon_{j}$. If $r_{m,n}$ RBs are assigned by AP $m$ to user $n$ for the first transmission, the bandwidth consumption can be calculated as $B_{m,n} = \alpha \beta r_{m,n} S_{cs} / \beta_{s}$, where $S_{cs}$ is the subcarrier spacing related to numerology and $\beta_{s}$ is the number of data symbol in one slot. Therefore, with the service rate $\omega_{m, n} = r_{m,n} \psi_{j} / T^{\text{s}}$, the effective capacity of user $n$ in \eqref{EC0} can be reinterpreted as
	\begin{equation}\label{sum of service rate}
		\begin{aligned}
			EC_{n}\left( \theta_{n} \right) 
			&= -\frac{1}{\theta_{n}T^{\text{s}}} \ln{\mathbb{E}_{\gamma}\left\{ \text{exp}\left(-\theta_{n}T^{\text{s}}\omega_{n}\right) \right\}} \\
			&= -\frac{1}{\theta_{n}T^{\text{s}}} \ln \mathbb{E}_{\gamma}\left\{ \text{exp}\left(-\theta_{n} \psi \sum_{m \in \mathcal{M}}  \!a_{m,n} r_{m,n}\right) \right\}.
		\end{aligned}
	\end{equation}
	
	For each time slot, it is assumed that the error block is retransmitted with the same AP and parameters used for the initial transmission. Hence, for the new data transmitted by AP $m$ to user $n$ within a slot, the number of RBs required for the $x$th transmission can be computed as $r_{m,n}^x = r_{m,n} (\bar{P}_{m,n})^{x-1}$. Finally, the average number of RBs consumed by AP $m$ for user $n$ in cumulative $X_n$ transmissions can be expressed as
	\begin{equation}\label{sum resource of single user}
		\begin{aligned}
			R_{m, n} & = r_{m, n}\! \cdot\! \sum_{x=0}^{X_{n}-1} \left(\bar{P}_{m, n}\right) ^x \\
			& = r_{m,n}\!\cdot\!\frac{1 - \left(\bar{P}_{m, n}\right) ^{X_n}}{1-\bar{P}_{m, n}}.
		\end{aligned}
	\end{equation}
	The average RB number that AP $m$ consumes for data transfer in the $t$-th slot ($t$ is large enough to be stable) can be calculated as
	\begin{equation}\label{AP resource consume}
		\begin{aligned}
			R_m[t] & =\sum_{n \in \mathcal{N}} a_{m,n} r_{m, n}[t] \\
			& +\!\sum_{n \in \mathcal{N}} \!\!a_{m,n}\!\sum_{x=1}^{X_n-1}\! r_{m, n}\!\left[t\!-\!x \cdot T^{\text{RTT}}\right] \cdot \left(\bar{P}_{m, n}\right)^{X_n-x} \\
			& \approx \!\sum_{n \in \mathcal{N}}\!\!a_{m,n} r_{m, n}[t]\!+\!\!\sum_{n \in \mathcal{N}} \!\!a_{m,n}\!\! \sum_{x=1}^{X_n-1} r_{m, n}[t] \!\cdot\!\left(\bar{P}_{m, n}\right)^x \\
			& =\sum_{n \in \mathcal{N}} \!\!a_{m,n} \sum_{x=0}^{X_n-1} r_{m, n}[t] \cdot \left(\bar{P}_{m, n}\right)^x \\
			& =\sum_{n \in \mathcal{N}} a_{m,n} R_{m, n}[t],
		\end{aligned}
	\end{equation}
	where $T^{\text{RTT}}$ denotes the static delay from one transmission to obtaining the feedback, including propagation delay and processing delay. The approximation is reasonable due to similar RB consumption in close slots and less amount of retransmissions. In summary, the total bandwidth consumed by the C-RAN system in slot $t$ can be calculated as
	\begin{equation}
		B[t] = \frac{\alpha \beta S_{cs}}{\beta_{s}} \sum_{m \in \mathcal{M}} \sum_{n \in \mathcal{N}} a_{m,n} R_{m, n}[t].
	\end{equation}
	The service delay due to $X_{n}$ transmission can be expressed as
	\begin{equation}\label{Dtr}
		D_{n}^{\text{s}}\left( X_{n} \right) = X_{n} \cdot \left( T^{\text{s}} + T^{\text{RTT}} \right).
	\end{equation}
	
	We aim to optimize the bandwidth efficiency by parameter configurations in each AP and proper UA decision under the constraints of delay and loss QoS requirements. Then the problem can be formulated as
		\begin{subequations}\label{Prob_Form}
			\begin{align}
				&\min_{^{\left\{r_{m,n} \right\}, \left\{ \theta_{n}\right\}, \left\{ a_{m,n}\right\}}_{\left\{\rho_{m,n}\right\}, \left\{X_{n}\right\}}}  ~
				\sum_{m \in \mathcal{M}}\sum_{n \in \mathcal{N}}\!\! a_{m,n} R_{m, n} \\ 
				&\text { s.t. } \quad \Pr\left\{ D_{n}^{\text{q}} > D_{n}^{\text{q,th}} \right\} \leq \varepsilon_{n}, \quad \forall n \in \mathcal{N}  \label{Constraint: queueLoss} \\
				&\quad\quad\quad EC_{n}\left( \theta_{n} \right) = \lambda_n,\quad  \forall n \in \mathcal{N} \label{Constraint: EC} \\ 
				&\quad\quad\quad D_{n}^{\text{s}} + D_{n}^{\text{q,th}} \leq D_{n}^{\text{th}}, \quad \forall n \in \mathcal{N} \label{Constraint: delay} \\
				&\quad\quad\quad \left( \bar{P}_{m,n} \right)^{X_{n}} \leq \varepsilon_{0}, \quad \forall n \in \mathcal{N},\forall m \in \mathcal{M} \label{Constraint: transLoss}\\
				&\quad\quad\quad \sum_{m \in \mathcal{M}}a_{m,n} \leq 1, \quad \forall n \in \mathcal{N} \label{Constraint:1UE1AP}\\
				&\quad\quad\quad \sum_{n \in \mathcal{N}}a_{m,n} \geq 1, \quad \forall m \in \mathcal{M} \label{Constraint:1APnUE}
			\end{align}
		\end{subequations}
	where \eqref{Constraint:1UE1AP} and \eqref{Constraint:1APnUE} are UA constraints implying that each user can be served by at most one AP and each AP serves at least one user, on the assumption that $N \geq M$. It is worth noting that the assumptions in \eqref{Constraint:1UE1AP} and \eqref{Constraint:1APnUE} controls the complexity of the cross-layer optimization problem, which will be further elaborated upon in subsequent sections of this paper. Based on effective capacity theory, the constraints \eqref{Constraint: queueLoss} - \eqref{Constraint: delay} implement statistical delay and loss QoS provision for applications with different source rate requirements. The \eqref{Constraint: transLoss} represents the BLER constraint of successful decoding, concurrently exerting a direct influence upon the transmission latency by means of the transmission count.

	\section{Problem Solution}\label{ProbSolu}
	The problem \eqref{Prob_Form} is a mixed-integer non-linear programming problem, indicating that its direct solution is typically challenging. Due to the fact that the optimal parameters are determined solely by the user with specified QoS requirements and the selected AP, the problem described in \eqref{Prob_Form} can be decomposed into two sub-problems, i.e., the TPD problem and UA problem, which can be addressed in sequence. First, we design the optimal transmission parameters for each potential AP-user pair in Sec. \ref{TPD Prob} to minimize the consumption of bandwidth resources while ensuring the QoS requirements of the users. Subsequently, we optimize the UA pattern based on the UA constraints and the optimal TPD for all possible pairs in Sec. \ref{UA Prob}.
	
	\subsection{Parameter Configuration for Each AP-User Pair} \label{TPD Prob}
	Consider a particular AP-user pair of the $m$th AP ($m \in \mathcal{M}$) and the $n$th user ($n \in \mathcal{N}$), which implies that the number of allocated RBs and the maximum tolerable queuing delay of user $n$ only depends on the parameter configuration of AP $m$. We first search for the optimum parameters configuration for users across all potential APs, so that the AP consumes the minimum amount of RBs while concurrently adhering to the constraints of the source rate and QoS. This forms sub-problem $\mathrm{\mathcal{P}1}$, the optimal solution of which is also the optimal parameter configuration after the UA algorithm has made its decision on the APs. Sub-problem $\mathrm{\mathcal{P}1}$ can be represented as
		\begin{subequations}\label{Prob_AP_para_conffig}
			\begin{align}
				\mathrm{\mathcal{P}1}:
				&\min_{^{\left\{r_{m,n} \right\}, \left\{ \theta_{m,n}\right\}}_{\left\{\rho_{m,n}\right\}, \left\{X_{m,n}\right\}}} ~ R_{m, n}\left(r_{m,n}, \theta_{m,n}, \rho_{m,n}, X_{m,n}\right) \\ 
				&\text { s.t. } \frac{\lambda_{m,n} T^{\text{s}}}{\mathbb{E}_{\gamma}\!\left[r_{m,n} \psi \right]} e^{-\theta_{m,n} \lambda_{m,n} D_{m,n}^{\text{q,th}}} \leq \varepsilon_{n} \label{Constraint: P1queueLoss} \\
				&\quad\quad -\frac{1}{\theta_{m,n} T^{\text{s}}} \ln \mathbb{E}_{\gamma}\left\{e^{-\theta_{m,n} r_{m,n} \psi}\right\} = \lambda_{m,n} \label{Constraint: P1EC} \\ 
				&\quad\quad D_{m,n}^{\text{s}} + D_{m,n}^{\text{q,th}} \leq D_{n}^{\text{th}} \label{Constraint: P1delay}\\
				&\quad\quad \left( \bar{P}_{m,n} \right)^{X_{m,n}} \leq \varepsilon_{0}. \label{Constraint: P1transLoss}
			\end{align}
		\end{subequations}
	It is worth noting that the utilization of subscripts $\left(m,n\right)$ instead of $\left(n\right)$ in this section serves the sole purpose of indicating that the parameter configuration for user $n$ remains valid exclusively within the context of the $m$th AP. Finally, we obtain an nonlinear programming problem with optimal solution $\left\{r_{m,n}^{\ast}, \theta_{m,n}^{\ast} , \rho_{m,n}^{\ast} ,X_{m,n}^{\ast}\right\}$.
	
	On the one hand, with an increased service latency $D_{m,n}^{\text{s}}$, a greater number of transmissions can be employed, thereby enabling the establishment of a larger BER threshold, as outlined by Eq. \eqref{Dtr} and the \textbf{Theorem \ref{thm1}} below. On the other hand, a relaxed queuing delay threshold $D_{m,n}^{\text{q,th}}$ can alleviate the impact of RB resources on LVP constraints, in accordance with Eq. \eqref{Constraint: P1queueLoss}. Both avenues contribute to a reduction in RB consumption. Consequently, it can be deduced that the optimal solution is assured when the condition set forth by Eq. \eqref{Constraint: P1delay} are satisfied as equality constraint. Furthermore, we obtain from \cite[Lemma 2, Theorem 1]{guo2019resource} that if the problem $\mathrm{\mathcal{P}1}$ is feasible, the inequality expressed as
	\begin{equation}\label{optimal EC Constraint}
		\lambda_{m,n}T^{\text{s}} < \mathbb{E}_{\gamma} \left\{r_{m,n}^{\ast}\psi\left(\rho_{m, n}^{\ast}\right)\right\} < \frac{\lambda_{m,n}T^{\text{s}}}{\varepsilon_{n}}
	\end{equation}
	holds in the optimal solution and constraint \eqref{Constraint: P1queueLoss} satisfy equality. Therefore, We can obtain the relational expression of $\theta_{m,n}$ with respect to $r_{m,n}$, $\rho_{m, n}$ and $X_{m,n}$ by limiting \eqref{Constraint: P1queueLoss} and \eqref{Constraint: P1delay} to equality, which can be calculated as
	\begin{equation}\label{theta calculation}
		\begin{aligned}
			\theta_{m,n} 
			&\overset{\eqref{Constraint: P1queueLoss}}{=} - \frac{1}{\lambda_{m,n} D_{m,n}^{\text{q,th}}} \ln\left\{ \frac{\varepsilon_{n}\mathbb{E}_{\gamma}\left[r_{m,n}\psi\right]}{\lambda_{m,n}T^{\text{s}}} \right\} \\
			&\overset{\eqref{Constraint: P1delay}}{=}\!\!- \frac{1}{\lambda_{m,n}\! \left[ D_{n}^{\text{th}} \!-\!\! X_{m,n}\!\cdot\!\left(T^{\text{s}}\!+\!T^{\text{RTT}}\right) \right]}\! \ln\!\left\{\! \frac{\varepsilon_{n}\mathbb{E}_{\gamma}\left[r_{m,n}\psi\right]}{\lambda_{m,n}T^{\text{s}}}\! \right\}.
		\end{aligned}
	\end{equation}
	Due to the finite set of integer variable values, $\mathrm{\mathcal{P}1}$ can be reformulated into a problem dependent upon $\theta_{m,n}$ and $\rho_{m, n}$ by prescribing specific values for $X_{m,n}$ and $r_{m,n}$. Thus, with fixed $r_{m,n}$ and $X_{m,n}$, as well as $F_{m,n}^{\text{ec}}\left(\rho_{m,n}\right)$ defined as
	\begin{equation}\label{F_mn}
		F_{m,n}^{\text{ec}}\left(\rho_{m,n}\right) = -\frac{1}{\theta_{m,n} T^{\text{s}}} \ln \mathbb{E}_{\gamma}\left\{e^{-\theta_{m,n} r_{m,n} \psi}\right\},
	\end{equation}
	the problem $\mathrm{\mathcal{P}1}$ can be transformed into a BER threshold configuration problem that is represented as
		\begin{subequations}\label{Prob_BER_conffig}
			\begin{align}
				\mathrm{\mathcal{P}2}:
				\min_{\rho_{m,n}} ~~&R_{m, n}\left(\rho_{m,n}\right) \\ 
				\text { s.t. } ~~ &F_{m,n}^{\text{ec}}\left(\rho_{m,n}\right) = \lambda_{m,n} \label{Constraint: P2EC} \\ 
				&\lambda_{m,n}T^{\text{s}} < \mathbb{E}_{\gamma} \left[r_{m,n}\psi\right] < \frac{\lambda_{m,n}T^{\text{s}}}{\varepsilon_{n}} \label{Constraint: P2ET} \\
				& \left[\bar{P}_{m,n}\left(\rho_{m, n}\right)\right]^{X_{m,n}} \leq \varepsilon_{0} \label{Constraint: P2transLoss}\\
				& R_{m, n}\left(\rho_{m,n}\right) \leq R^{\text{s}}, \label{Constraint: P2resources}
			\end{align}
		\end{subequations}
	where $R^{\text{s}}$ denotes the total number of RBs in one slot and the constraint \eqref{Constraint: P2ET} obtained from \eqref{optimal EC Constraint} further limits $\theta_{m,n} > 0$.
	
	According to the definition of spectrum efficiency $\upsilon_{j}$ and average BLER $\bar{P}$, we present two properties of the optimal solution of problem  $\mathrm{\mathcal{P}2}$, in what follows by \textbf{Theorem \ref{thm1}} and \textbf{Theorem \ref{thm2}}, proved in Appendix A and Appendix B respectively.
	
	\begin{thm1}\label{thm1}
		Based on the calculation of BER in \eqref{actual BER} and the definition of spectrum efficiency \cite{3gpp_technical_38214} for each MCS mode in NR, the average BLER $\bar{P}\left(\rho\right)$ is monotonically increasing with BER threshold $\rho$ for $\rho > 0$ when the distribution and average quality of channel are determined.
	\end{thm1}
	
	\begin{thm1}\label{thm2}
		If constraint \eqref{Constraint: P2ET} holds, the $F_{m,n}^{\text{ec}}\left(\rho_{m,n}\right)$ is monotonically increasing with $\rho_{m,n}$ for $\rho_{m,n}>0$.
	\end{thm1}
	
	From the proof of \textbf{Theorem \ref{thm2}}, we also know that the expectation of the amount of information per RB $\mathbb{E}_{\gamma}\left[\psi\right]$ can be calculated as
	\begin{equation}\label{E_psi}
		\begin{aligned}
			\mathbb{E}_{\gamma}\left[\psi\right]
			&=\frac{1}{P_{T}} \sum_{j=0}^{J} \int_{\gamma_{j}\left(\rho\right)}^{\gamma_{j+1}\left(\rho\right)} \psi_{j}\cdot p_{\gamma}\left(\gamma\right) d\gamma \\
			&=\frac{1}{P_{T}} \sum_{j=0}^{J} \int_{\gamma_{j}\left(\rho\right)}^{\gamma_{j+1}\left(\rho\right)} \alpha\cdot\beta\cdot\upsilon_{j}\cdot p_{\gamma}\left(\gamma\right) d\gamma,
		\end{aligned}
	\end{equation}
	which is monotonically increasing with $\rho_{m, n}$. Therefore, we can find the optimal solution $\rho_{m, n}^{\ast}$ corresponding to each AP-user pair by binary search. The $\theta_{m,n}$ is first calculated by \eqref{theta calculation}, where $\mathbb{E}_{\gamma}\left[\psi\right]$ can be obtained by \eqref{E_psi}, and the effective capacity in \eqref{F_mn} can be rewritten as
	\begin{equation}
		\begin{aligned}
			&F_{m,n}^{\text{ec}}\left(\rho_{m,n}\right) = -\frac{1}{\theta_{m,n} T^{\text{s}}} \ln \mathbb{E}_{\gamma}\left[e^{-\theta_{m,n} r_{m,n} \psi_{j}}\right] \\
			&=\!-\frac{1}{\theta_{m,n} T^{\text{s}}}\! \ln\! \left\{\!\frac{1}{P_{T}}\! \sum_{j=0}^{J}\! \int_{\gamma_{j}\left(\rho_{m\!, n}\right)}^{\gamma_{j\!+\!1}\left(\rho_{m\!, n}\right)}\!\! \left[e^{-\theta_{m\!,n} r_{m\!,n} \psi_{j}}\right]\! p_{\gamma}\!\!\left(\gamma\right)\! d\gamma\right\}.
		\end{aligned}
	\end{equation}
	For each AP-user pair, the optimal solution to problem $\mathrm{\mathcal{P}1}$  can be obtained by solving problem $\mathrm{\mathcal{P}2}$ for all feasible transmission count and RB quantities. The proposed comprehensive optimization solution is summarized in \textbf{Algorithm \ref{alg:P1bisearch}}, encompassing the procedure for solving problem $\mathrm{\mathcal{P}2}$. It is worth noting that a necessary condition to ensure the feasibility of problem $\mathrm{\mathcal{P}2}$ is $F_{m,n}^{\text{ec}}\left(\rho_{m, n}^{\text{max}}\right) \geq \lambda_{m,n}$, as $F_{m,n}^{\text{ec}}\left(\rho_{m, n}\right)$ increases monotonically with $\rho_{m, n}$.
	
	\begin{algorithm}[t]
		\caption{Solution Procedure for Problem $\mathrm{\mathcal{P}1}$ using Binary Search}\label{alg:P1bisearch}
		\KwIn{Minimum and maximum value $\odot^{\text{min}}$, $\odot^{\text{max}}$ of $\rho_{m,n}$ and $X_{m,n}$, Average SNR $\bar{\gamma}_{m,n}$,
			
			\quad \quad \quad Traffic arrival rate $\lambda_{m, n}$, Total latency $D_{n}^{\text{th}}$,
			
			\quad \quad \quad Decoding BLER threshold $\varepsilon_{0}$, LVP $\varepsilon_{n}$,
			
			\quad \quad \quad Total number of RBs for system $R^{\text{s}}$}
		\KwOut{\emph{The optimal parameter configuration $\left\{r_{m,n}^{\ast}, \rho_{m,n}^{\ast}, X_{m,n}^{\ast}\right\}$ for AP $m$ and user $n$}}
		
		Let the optimal value $R_{m\!,n}^{\ast} = +\infty$, 
		
		Set error tolerance $\epsilon$;
		
		\For{$X_{m,n}=1 \textrm{ to } X_{m,n}^{\text{max}}$}{
			Calculate $D_{m,n}^{\text{q,th}}$ by \eqref{Dtr} and \eqref{Constraint: P1delay};
			
			\For{$r_{m,n}=1 \textrm{ to } R^{\text{s}}$}{
				\If{$r_{m,n} > R_{m\!,n}^{\ast}$}{break;}
				\If{$F_{m,n}^{\text{ec}}\left(\rho^{\text{max}}\right) \geq \lambda_{m, n}$}{
					Set $\rho_{m, n}^{\text{min}}=\rho^{\text{min}}$ and $\rho_{m, n}^{\text{max}}=\rho^{\text{max}}$;
					
					\Repeat{$| \rho_{m, n}^{\text{max}}-\rho_{m, n}^{\text{min}} | \leq \epsilon$}{
						$\rho_{m, n} = \left( \rho_{m, n}^{\text{max}} + \rho_{m, n}^{\text{min}} \right)/2$ ;\\
						
						\eIf{$F_{m,n}^{\text{ec}}\left( \rho_{m, n} \right) < \lambda_{m, n}$}{
							$\rho_{m, n}^{\text{min}} = \rho_{m, n}$;
						}{
							$\rho_{m, n}^{\text{max}} = \rho_{m, n}$;
						}
					}
					
					\If{constraints \eqref{Constraint: P2ET}-\eqref{Constraint: P2resources} are satisfied \& $R_{m\!,n}^{\ast} > R_{m\!,n}\left(r_{m\!, n}, \rho_{m\!,n},  X_{m\!,n}\right)$}{
						$R_{m\!,n}^{\ast} = R_{m\!,n}\left(r_{m\!, n}, \rho_{m\!,n},  X_{m\!,n}\right)$;
						
						$\left\{\!r_{m,n}^{\ast},\! \rho_{m,n}^{\ast},\! X_{m,n}^{\ast}\!\right\} \!=\! \left\{\!r_{m\!, n}, \!\rho_{m\!,n},\!  X_{m\!,n}\!\right\}$;
						
					}
				}	
			}
		}
		\Return{$R_{m\!,n}^{\ast}$, and $\left\{\!r_{m,n}^{\ast},\! \rho_{m,n}^{\ast},\! X_{m,n}^{\ast}\!\right\}$ if it exists.}
	\end{algorithm}
	
	\subsection{UA Decision}\label{UA Prob}
	In this section, the UA decisions for users are studied based on the minimum bandwidth consumption $R_{m,n}^{\ast}$ for each AP-user pair obtained from section \ref{TPD Prob}, with the aim of minimizing the bandwidth cost of the system. Given fixed transmission parameters, the problem can be formulated as an asymmetric multi-assignment problem as follows
	\begin{subequations}\label{P:UA optimization}
		\begin{align}
			\mathrm{\mathcal{P}3}:
			\min_{a_{m,n}}
			&\sum_{m \in \mathcal{M}}\sum_{n \in \mathcal{N}} a_{m,n} R_{m,n}^{\ast} \\ 
			\text { s.t. }
			&\sum_{m \in \mathcal{M}}a_{m,n} = 1, \quad \forall n \in \mathcal{N} \label{UAConstraint: 1UE1AP} \\ 
			&\sum_{n \in \mathcal{N}}a_{m,n} \geq 1, \quad \forall m \in \mathcal{M} \label{UAConstraint: 1APnUE} \\ 
			&a_{m,n} \in \left\{0,1\right\}, \quad \forall (m,n) \in \mathcal{A},\label{UAConstraint: amn}
		\end{align}
	\end{subequations}
	where $\mathcal{A}$ represents the set of all AP-user pairs and the condition in \eqref{Constraint:1UE1AP} is converted to an equality constraint due to positive resource rate. Bandwidth consumption $R_{m,n}^{\ast}$ can be viewed as the cost paid by AP $m$ to satisfy the QoS constraints of user $n$ at the source rate $\lambda_{n}$, thus the objective function is the total system cost. We define $\mathcal{S}$ as a subset of $\mathcal{A}$, consisting of paired AP-user pairs $(m,n)$. Due to the UA constraints \eqref{UAConstraint: 1UE1AP} and \eqref{UAConstraint: 1APnUE}, each AP $m$ is a part of at least one pair $(m,n) \in \mathcal{S}$ and each user $n$ is a part of only one pair $(m,n) \in \mathcal{S}$. By setting $a_{m,n}=1$ if $(m,n) \in \mathcal{S}$ and $a_{m,n}=0$ otherwise, we can obtain a feasible assignment $\mathcal{S}$ of problem $\mathrm{\mathcal{P}3}$. Based on network optimization theory \cite{zhao2021joint,athanasiou2013auction}, we can transform the problem $\mathrm{\mathcal{P}3}$ into a minimum cost flow problem by introducing a supersource node $e$ connected to each AP, which can be represented as
	\begin{subequations}\label{P:min cost}
		\begin{align}
			\min_{a_{m,n}}
			&\sum_{m \in \mathcal{M}}\sum_{n \in \mathcal{N}} R_{m,n}^{\ast}a_{m,n} \\ 
			\text { s.t. }
			&\sum_{n \in \mathcal{N}}a_{m,n} - a_{e,m} = 1,  \forall m \in \mathcal{M} \label{MCConstraint: AP flow 1 unit} \\ 
			&\sum_{m \in \mathcal{M}}a_{e,m} = N - M, a_{e,m}\geq 0, \forall m \in \mathcal{M} \label{MCConstraint: e flow nm unit} \\ 
			&\sum_{m \in \mathcal{M}} a_{m,n} = 1, \forall n \in \mathcal{N}, a_{m,n}\geq 0, \forall (m,n) \in \mathcal{A},\label{MCConstraint: 1UE1AP}
		\end{align}
	\end{subequations}
	where $a_{m,n}$ is extended to include the supersource node $e$ and the constraint \eqref{MCConstraint: AP flow 1 unit} implies that each AP generates one unit flow. Restriction \eqref{MCConstraint: e flow nm unit} means that the number of streams generated by the supersource node and AP is equal to the number of users, and \eqref{MCConstraint: 1UE1AP} restricts each user to request one unit stream. The optimal solution of problem \eqref{P:min cost} is the same to the initial asymmetric multi-assignment problem \eqref{P:UA optimization}.
	
	{\setlength\abovedisplayskip{1pt}
		\setlength\belowdisplayskip{1pt}
		\begin{algorithm}[t]
			\caption{\mbox{Auction-based Decision for AP Selection}} \label{Auction}
			\SetKwInOut{KwIn}{Input}
			\nlset{1}\KwIn{$\mathbf{R}^{\ast}$, $\left\{\mathcal{M}\right\}_{n=1}^{N}$, $\left\{\mathcal{N}\right\}_{m=1}^{M}$, 
				$\mathcal{M}^{\circ}\!=\!\mathcal{M}$, $\mathcal{N}^{\circ}\!=\!\mathcal{N}$,
				Init of $\mathcal{S},(\pi,p), \epsilon$ and $\mu$.}
			\textbf{Ensure:}{\ $\mathcal{S} \text{ and } (\pi,p)\ \text{satisfy }\eqref{eCc-1}\text{ and } \eqref{eCc-2};$}\ \ \ \ 
			Repeat following cycles until $\mathcal{M}^{\circ}\!=\! \mathcal{N}^{\circ}\!=\!\varnothing$;
				
				\While{$\mathcal{M}^{\circ} \neq \varnothing$}{
					\For{$m \in \mathcal{M}^{\circ}$}{
						find its best user $n_m$ that\\
						$n_m = \mathop{\arg \min}_{n \in \mathcal{N}} \left\{R_{m,n}^{\ast} + p_n\right\}$.\\
						$\xi_m = \min_{n \in \mathcal{N}} \left\{R_{m,n}^{\ast} + p_n\right\}$;\\
						$\zeta_m = \min_{n \in \mathcal{N}, n \neq n_m} \left\{R_{m,n}^{\ast} + p_n\right\}$;\\
						\If{$n_m$ is the only user in $\mathcal{N}$}{
							$\zeta_m  \rightarrow -\infty$;\\}
						$ b_{m,n_m} = p_{n_m} \!+\! \xi_m \!-\! \zeta_m \!+\! \epsilon \!=\! -R_{m,n_m}^{\ast} \!-\! \zeta_m \!+\! \epsilon$;\\
					}
					
					\For{$n \in \mathcal{N} \ \&\  \mathcal{M}^{\text{b}}\neq \varnothing$}{
						$\mathcal{M}^{\text{b}}$ is set of APs from which user $ n $ received a bid during bidding.\\
						$\mathcal{S}=\mathcal{S} \backslash \{(m', n)\}$;
						$\mathcal{S}=\mathcal{S} \cup \{(m_n,\! n)\}$ with $ m_n \!\!=\!\! \mathop{\arg\max}_{m \in \mathcal{M}^{\text{b}}} \!b_{m,n} $;\\
						Update $\mathcal{M}^{\circ}$ and $\mathcal{N}^{\circ}$;\\
						\mbox{$p_n \!:=\! \max_{m \in \mathcal{M}^{\text{b}}} b_{m,n}$; $\pi_{m_n}\!\!=\!-(R_{m_n,n}^{\ast}\!\!+\!p_n)$.}
					}
				}
				$\mu = \mathop{\max}_{m = 1, \cdots, M} \pi_m$;
				
				\While{$\mathcal{N}^{\circ} \neq \varnothing$}{
					Select user $n \!\in\! \mathcal{N}^{\circ}$, find its best AP $m_n$ that\\
					$m_n = \mathop{\arg\min}_{m \in \mathcal{M}} \left\{R_{m,n}^{\ast} + \pi_m\right\}$.\\ 
					$ \chi_n = \min_{m \in \mathcal{M}} \left\{R_{m,n}^{\ast} + \pi_m\right\}$,\\ 
					$\zeta_n = \min_{m \in \mathcal{M}, m \neq m_n} \left\{R_{m,n}^{\ast}+ \pi_m\right\}$;\\
					\If{$m_n$ is the only AP in $\mathcal{M}$}
					{
						$\zeta_n \rightarrow -\infty$;
					}
					$\mathcal{S}=\mathcal{S} \cup \{(m_n,\! n)\}$;\\
					$\delta = \min\left\{\mu - \pi_{m_n}, \chi_n - \zeta_n + \epsilon \right\}$;\\
					$\pi_{m_n} = \pi_{m_n} + \delta, \ p_n = \chi_n - \delta $;\\
					\If{$\delta > 0$}{
						$\mathcal{S}=\mathcal{S} \backslash \{(m_n,\! n')\}$, where $n'$ was assigned to $m_n$ under $\mathcal{S}$;
					}
					Update $\mathcal{M}^{\circ}$ and $\mathcal{N}^{\circ}$.
				}
				\KwOut{A feasible optimal assignment $\mathcal{S}$.}
			\end{algorithm}
		}
		
		By leveraging duality theory, we can express the duality problem of the minimum cost flow problem  \eqref{P:min cost} as follows:
		\begin{subequations}\label{P:min cost dual}
			\begin{align}
				\min_{\pi_m,p_n,\mu} ~
				&\sum_{m \in \mathcal{M}} \pi_m + \sum_{n \in \mathcal{N}} p_n + (N-M)\mu \\
				\text{ s.t. }~
				&\pi_{m} + p_n \geq -R_{m,n}^{\ast}, \quad \forall (m,n) \in \mathcal{A} \\
				&\mu \geq \pi_m, \quad \forall m \in \mathcal{M} 
			\end{align}
		\end{subequations}
		where $-\pi_m$, $\mu$ and $p_n$ are Lagrange multipliers associated with \eqref{MCConstraint: AP flow 1 unit}, \eqref{MCConstraint: e flow nm unit}, and \eqref{MCConstraint: 1UE1AP}, respectively. We denote $-\pi_m$ as the price of each AP $m$, $\mu$ as the price of the supersource node $e$, and $p_n$ as the price of each user $n$. The optimal solution to problem \eqref{P:UA optimization} can be derived from the optimal solution to problem \eqref{P:min cost dual}. Then the \emph{$\epsilon$-complementary slackness ($\epsilon-$CS)} is introduced to solve the multi-assignment problem. By introducing a positive scalar $\epsilon$, we declare that the assignment $\mathcal{S}$ and the binary pair $(\pi,p)$ satisfy the \emph{$\epsilon-$CS} condition if
		\begin{subequations}\label{e-CS condition}
			\begin{align}
				&\pi_m +p_n \geq -R_{m,n}^{\ast} - \epsilon,\forall (m,n) \in \mathcal{A} \label{eCc-1}\\
				&\pi_m + p_n = -R_{m,n}^{\ast}, \forall (m,n) \in \mathcal{S} \label{eCc-2}\\
				&\pi_m = \max_{k=1,...,M} \pi_k, \text{ if }m \text{  has multi-pairs }(m,n) \in \mathcal{S} \label{eCc-3}
			\end{align}
		\end{subequations}
		
		\begin{thm1}\label{thm3}
			Assume that the benefits $-R_{m,n}^{\ast}$ are integer. If \emph{$\epsilon-$CS} condition are satisfied by feasible assignment $\mathcal{S}$ and the binary pair $(\pi,p)$ for $\epsilon < 1/M$, then $\mathcal{S}$ is the optimal assignment for the initial asymmetric multi-assignment problem. 
		\end{thm1}
		
		\begin{proof}
			The proof naturally originates from \cite[Proposition 1]{zhao2021joint}, thus, it will not be elaborated in this paper.
		\end{proof}
		
		\begin{figure}[bt]
			\centering
			\includegraphics[scale=0.5]{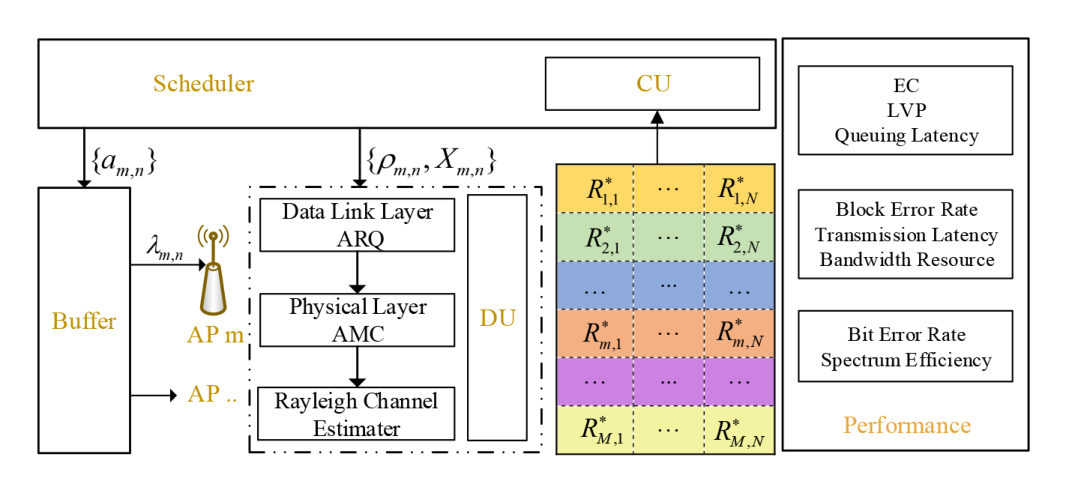}
			\caption{The system framework of the distributed TPD and UA for cross-layer optimization of C-RAN network.}
			\label{Fig:system framework}
		\end{figure}
		
		Based on the significance of the \emph{$\epsilon-$CS} condition stated in the \textbf{Theorem \ref{thm3}}, The auction-based algorithm is proposed to solve the asymmetric multi-assignment  problem, which is explained in \textbf{Algorithm \ref{Auction}}. The forward (line 3-20) and reverse auction (line 21-37) processes are iteratively executed until the sets of unallocated APs and users satisfy $\mathcal{M}^{\circ}\!=\! \mathcal{N}^{\circ}\!=\!\varnothing$. In the forward auction, each unassigned AP $m$ selects its best user $n_m$ based on the maximum profit $\min_{n \in \mathcal{N}} \left\{R_{m,n}^{\ast} + p_n\right\}$ and bid on it (line 4-13). The user $n$ that receives bids will be assigned to the AP $m_n$ with the highest bid, and the pair $(m_n,n)$ will be added to the assignment $\mathcal{S}$. Any existing association between the user $n$ and other AP will be cancelled and the values of $(\pi,p)$ are updated to satisfy the \emph{$\epsilon-$CS} condition (line 14-19). Reverse auction is executed after the maximum initial AP profit update. The unassigned user $n$ selects its best AP $m_n$ based on the maximum price $\mathop{\arg\min}_{m \in \mathcal{M}} \left\{R_{m,n}^{\ast} + \pi_m\right\}$ and adds the AP-user pair $(m_n,n)$ to the assignment $\mathcal{S}$ (line 23-30). The allocation needs to be adjusted further to meet \emph{$\epsilon-$CS} condition (line 31-36).
		
		\begin{figure*}[htbp]
			\centering
			\subfigtopskip=2pt 
			\subfigbottomskip=2pt 
			\subfigcapskip=-5pt 
			
			\subfigure[bandwidth consumption]{
				\label{Fig:modelVa_Band_lambda_1}
				\includegraphics[width=0.32\linewidth]{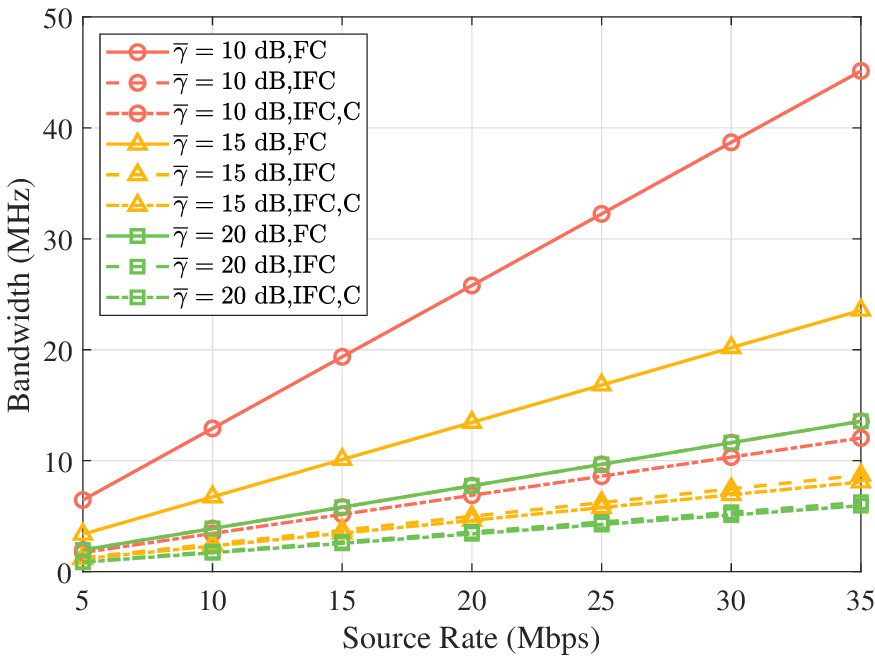}}
			\subfigure[optimal BER threshold]{
				\label{Fig:modelVa_rho_lambda_2}
				\includegraphics[width=0.32\linewidth]{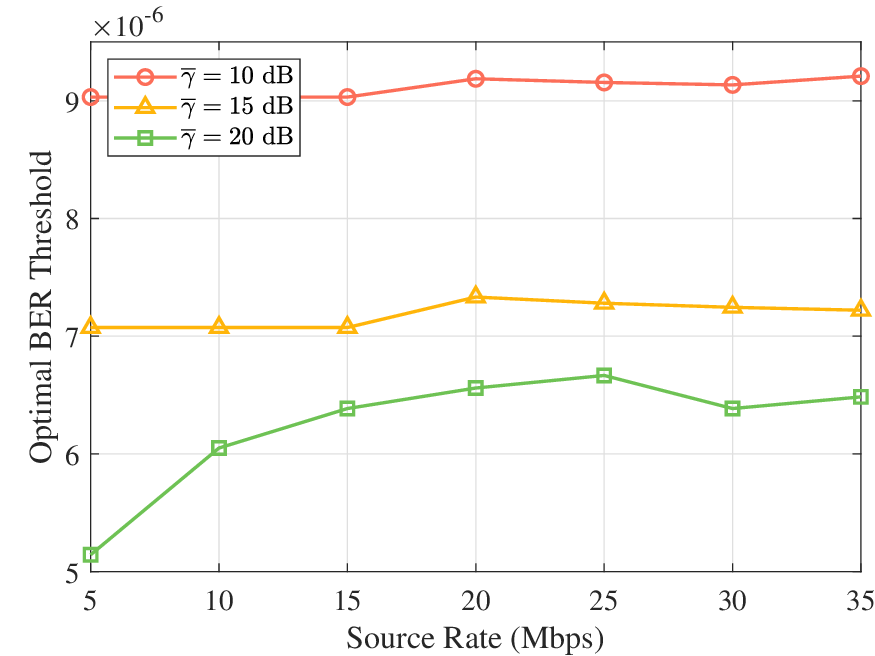}}
			\subfigure[BLER of practical FC system]{
				\label{Fig:modelVa_BLER_SNR_3}
				\includegraphics[width=0.32\linewidth]{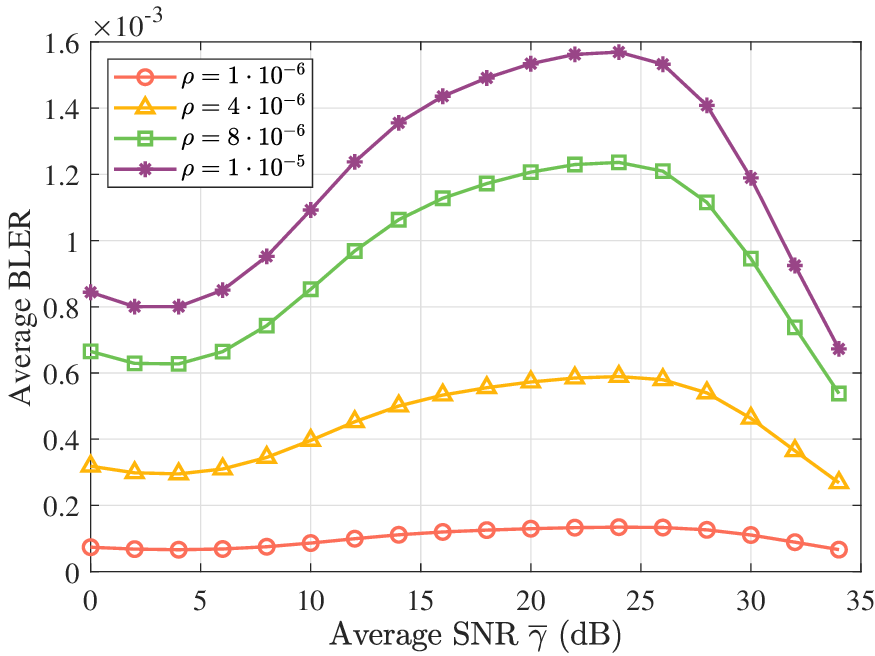}}
			
			\caption{Bandwidth consumption and optimal BER threshold selection varying with the source data rate under different average channel quality, where the total latency budget $D_n^{\text{th}}=6$ ms, the delay violation probability $\varepsilon_n =10^{-5}$ and the decoding BLER threshold $\varepsilon_0 = 10^{-3}$.}
			\label{Fig:modelVa1}
		\end{figure*}
		
		\subsection{Algorithm Deployment and Complexity Analysis}
		The deployment of the algorithm can be either centralized, leveraging the abundant computing resources provided by the cloud, or distributed, taking advantage of parallel processing at APs. As an illustration, we propose a distributed scheme to implement our proposed algorithm. The solution is based on the idea of multi-point collaborative processing, which is fully illustrated in Fig. \ref{Fig:system framework}. The calculation of transmission parameters is implemented in the distributed units (DUs) equipped in each AP, based on the average channel quality and QoS requirements of each user. The optimal transmission parameters and their corresponding minimum bandwidth consumption for serving different users in a slot will be transmitted to the scheduler through the backhaul link. The UA algorithm for each user determined by the CU of the scheduler through \textbf{Algorithm \ref{Auction}} is implemented in the buffer. It is worth noting that the parameter table needs to be updated only when the channel quality changes, which significantly reduce the performance requirements of AP and backhaul links.
		
		With the analysis above, the time complexity of \textbf{Algorithm \ref{alg:P1bisearch}} for each AP-user pair can be represented as $\mathcal{O}(X^{\text{max}} R^{\text{s}}\log{1/\epsilon})$. Therefore, the average complexity of \textbf{Algorithm \ref{alg:P1bisearch}} deployed on each AP is $\mathcal{O}(\bar{N} X^{\text{max}} R^{\text{s}}\log{1/\epsilon})$, where $\bar{N} =\frac{1}{M} \sum_{m=1}^{M}|\mathcal{N}|$ is the average number of times \textbf{Algorithm \ref{alg:P1bisearch}} is triggered at each AP. According to \cite{zhao2021joint}, the complexity of \textbf{Algorithm \ref{Auction}} executed by the scheduler is mainly determined by the value of $\epsilon$ and the greatest difference of the cost $\Delta=\max\limits_{(m,n)\in \mathcal{A}}R_{m,n}^{\ast}-\min\limits_{(m,n)\in \mathcal{A}}R_{m,n}^{\ast}$. The time complexity of forward and reverse auction can be expressed as $\lceil \Delta/\epsilon \rceil|\mathcal{A}|=\mathcal{O}(\lceil\Delta/\epsilon\rceil MN)$ and $\mathcal{O}(\lceil\Delta/\epsilon\rceil M(N-M))$, respectively, which shows that forward auction is dominant due to $N>N-M$. 
		
		\section{Performance Evaluation}\label{Performance E}
		In this section, extensive simulations and discussions are presented to evaluate the performance of the proposed optimal TPD algorithm based on the finite-length coding and the UA algorithm based on the auction (i.e., the proposed \textbf{Algorithm \ref{alg:P1bisearch}} and \textbf{Algorithm \ref{Auction}}, respectively) in the C-RAN system. Prior to conducting performance evaluations, we first examine the impact of flexible TPD on performance to verify the rationality and practicality of the model. Subsequently, ablation experiments and performance comparisons with baseline schemes are conducted to validate the effectiveness of our proposed algorithms for QoS guarantee and bandwidth saving.
		
		\subsection{Model Validation}\label{Sec: Model Validation}
		The coupling between transmission parameters and their relationship with performance requirements is elucidated to validate the rationality of finite-length encoding models and effective capacity models, which serve as the foundation of performance assessment. Taking into account a user $n$ being served by AP $m$, we generate a random channel realization with average channel quality $\bar{\gamma}$ to evaluate the bandwidth consumption obtained by \textbf{Algorithm \ref{alg:P1bisearch}} under various channel conditions for services with different source rates and specific QoS requirements. With the maximum total service delay limitation $D_{n}^{\text{th}}=6$ ms, the queue LVP $\varepsilon_n = 10^{-5}$, and the decoding BLER threshold $\varepsilon_{0}=10^{-3}$, the simulation results are depicted in Fig. \ref{Fig:modelVa1}. The line types represent bandwidth consumption obtained by Shannon channel capacity (IFC, C) for specific source rates, considering delay and LVP additionally (IFC) and actual finite-length coding system on top of the previous considerations (FC), respectively.

		The bandwidth consumption obtained by the Shannon capacity assumes that there exists a theoretically perfect modulation and coding scheme that enables error-free transmission. This results in the minimum bandwidth requirement and a relatively small increase when considering the delay requirement and LVP, as illustrated in Fig. \ref{Fig:modelVa_Band_lambda_1}. When the actual MCS is implemented considering BLER requirements, significantly increased bandwidth resources are consumed serving the same source rate, due to the low spectral efficiency affected by limited BER threshold. The situation becomes even worse when channel deteriorates because of the lower modulation order, i.e., the lower spectral efficiency, determined by the switching threshold and channel quality in the actual MCS. Furthermore, the average BLER increases with channel quality optimization when the channel SNR is in $[10,20]$ dB due to the characteristics of actual MCS, as illustrated in Fig. \ref{Fig:modelVa_BLER_SNR_3}, which results in lower BER threshold selection in Fig. \ref{Fig:modelVa_rho_lambda_2} with better channel quality. The above results verify the rationality of the finite length coding model and the effective capacity model.
		
		\begin{figure*}[tb]
			\centering
			\subfigtopskip=2pt 
			\subfigbottomskip=2pt 
			\subfigcapskip=-5pt 
			
			\subfigure[bandwidth consumption]{
				\label{Fig:modelVa_BDthGamma_vTheta_1}
				\includegraphics[width=0.32\linewidth]{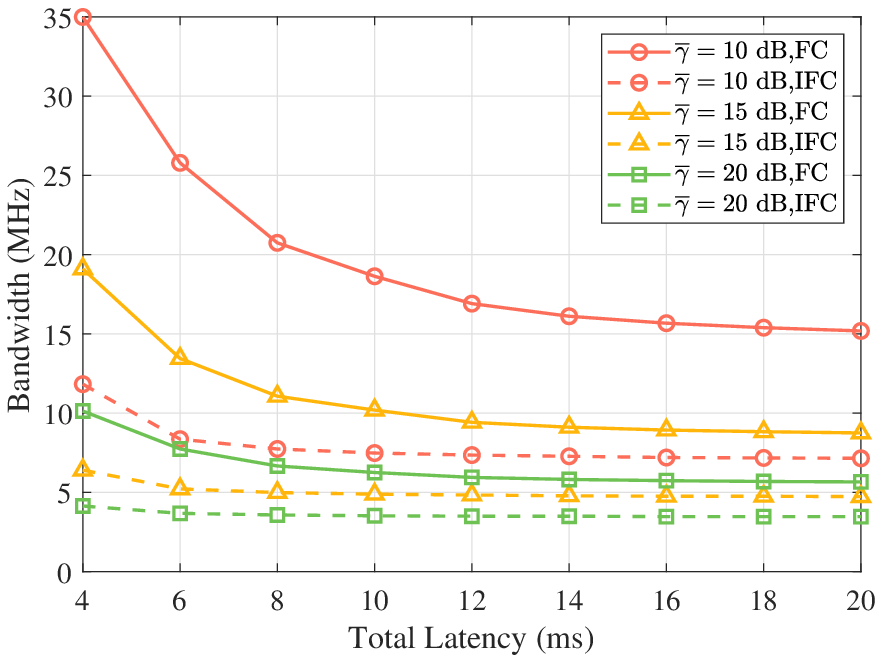}}
			\subfigure[optimal BER threshold and number of transfers]{
				\label{Fig:modelVa_RhoDthGamma_vTheta_2}
				\includegraphics[width=0.32\linewidth]{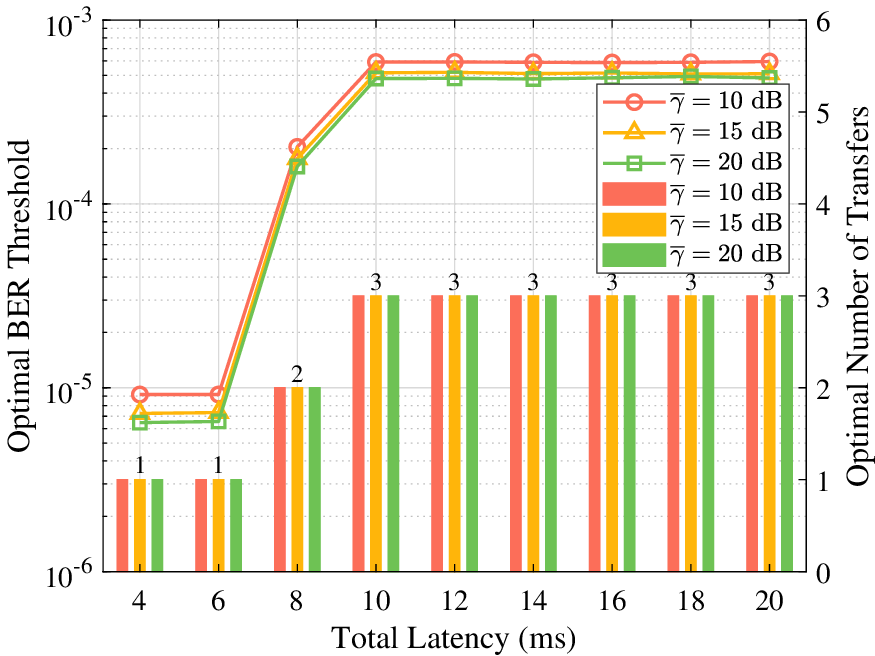}}
			\subfigure[Spectrum Efficiency of practical FC system]{
				\label{Fig:modelVa_Eff_BERth_3}
				\includegraphics[width=0.32\linewidth]{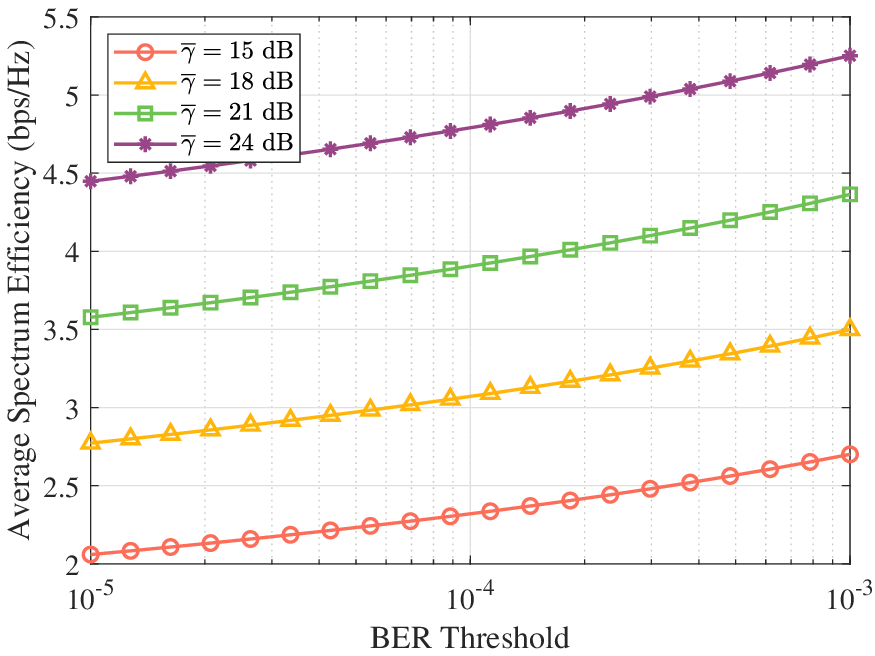}}
			\caption{Bandwidth consumption, the optimal BER threshold selection and the corresponding optimal number of transmissions varying with the total latency budget under different average channel quality, where the source data rate $\lambda=20$ Mbps, the LVP $\varepsilon_n =10^{-5}$ and the decoding BLER threshold $\varepsilon_0 = 10^{-3}$.}
			\label{Fig:modelVa2}
		\end{figure*}
		
		Fig. \ref{Fig:modelVa2} illustrates the bandwidth consumption and the optimal transmission parameters under various channel conditions, affected by different overall transmission delay constraints, when serving the same source rate with an identical LVP and decoding BLER threshold. In cases where the total latency is minimal ($D^{\text{th}}\leq 6$ ms), the scarcity of time resources is the primary concern. Providing services that can tolerate higher queuing delay results in greater bandwidth savings, surpassing the effect achieved by higher spectral efficiency due to increased BER thresholds caused by retransmissions. As the total latency constraint is moderately relaxed ($6 \leq D^{\text{th}}\leq 10$ ms), the more efficient bandwidth utilization brought about by the increase in BER threshold provides a greater contribution to bandwidth savings. Consequently, the maximum transmission count increases to adopt a higher BER threshold, thereby enhancing bandwidth efficiency at the expense of a reduced tolerable queuing delay. The enhancement effect of increasing the BER threshold on spectral efficiency is not always prominent but gradually diminishes, as illustrated in Fig. \ref{Fig:modelVa_Eff_BERth_3}. Therefore, when the total latency constraint is further relaxed ($D^{\text{th}}\geq 10$ ms), tolerating higher queuing latency once again becomes the primary contributor to our objective. Consequently, the decision to maintain a constant transmission count and BER threshold is made again.

\begin{figure}[htbp]
			\centering
			\subfigtopskip=2pt 
			\subfigbottomskip=2pt 
			\subfigcapskip=-5pt 
			
			\subfigure[bandwidth consumption]{
				\label{Fig:modelVa_BLVPGamma_vTheta_1}
				\includegraphics[width=0.75\linewidth]{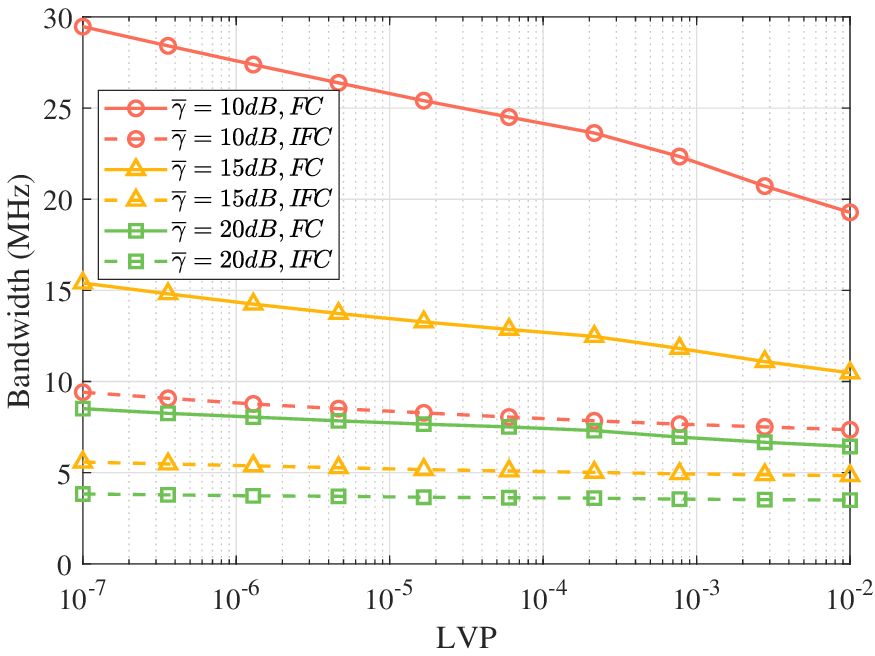}}
			\subfigure[optimal BER threshold and number of transfers]{
				\label{Fig:modelVa_RhoLVPGamma_vTheta_2}
				\includegraphics[width=0.82\linewidth]{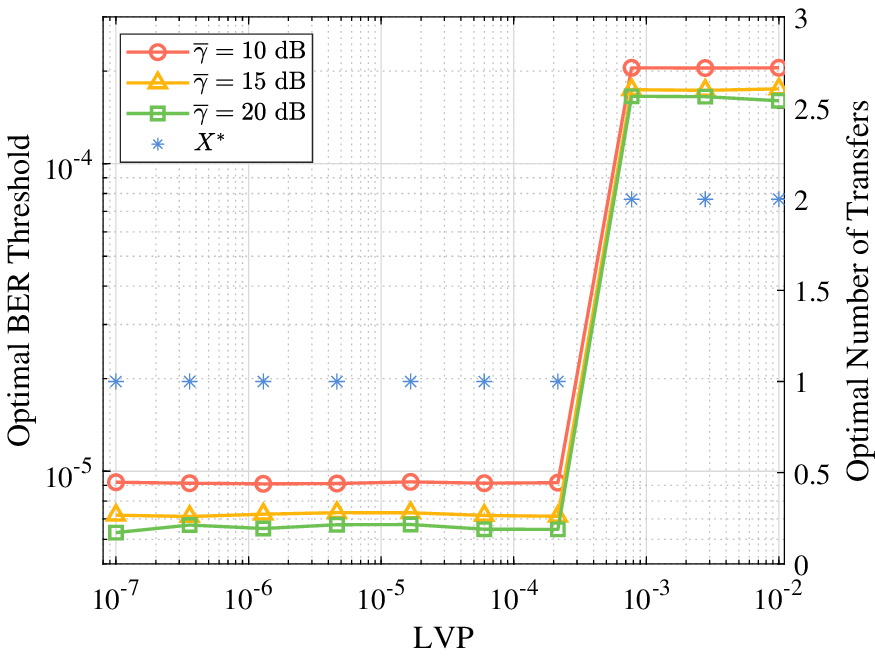}}
			\caption{Bandwidth consumption, the optimal BER threshold selection and the corresponding optimal number of transmissions varying with the LVP under different average channel quality, where the source data rate $\lambda=20$ Mbps, the total latency budget $D^{\text{th}}=6$ ms and the decoding BLER threshold $\varepsilon_0 = 10^{-3}$.}
			\label{Fig:modelVa3}
		\end{figure}

		Fig. \ref{Fig:modelVa3} illustrates the bandwidth consumption and optimal transmission parameters under various channel conditions, affected by different LVP constraints, when serving the same source rate with an identical overall delay budget and decoding BLER threshold. Consistent with our analysis, the required bandwidth monotonically decreases with LVP, and the difference in bandwidth demand for different LVPs is more pronounced under poorer channel conditions, as illustrated in Fig. \ref{Fig:modelVa_BLVPGamma_vTheta_1}. When services with strict LVP requirements ($\varepsilon_n \leq 2\cdot 10^{-4}$) are requested, fewer transmissions are adopted, resulting in a larger available queuing delay, which alleviates bandwidth consumption of the strict LVP. For services with relaxed LVP requirements, our algorithm transfers part of the queuing latency budget to the transmission latency budget, which enables a higher BER threshold to improve the spectral efficiency, as evidenced by the increased resource savings ($\varepsilon_n > 2\cdot 10^{-4}$) in Fig. \ref{Fig:modelVa_BLVPGamma_vTheta_1}. The reason is that a higher BER threshold saves more bandwidth than a larger queuing delay budget under more relaxed LVP constraints.
		
		\begin{figure}[tb]
			\centering
			\subfigtopskip=2pt 
			\subfigbottomskip=2pt 
			\subfigcapskip=-5pt 
			
			\subfigure[bandwidth consumption]{
				\label{Fig:modelVa_BLVPBER_vTheta_1}
				\includegraphics[width=0.8\linewidth]{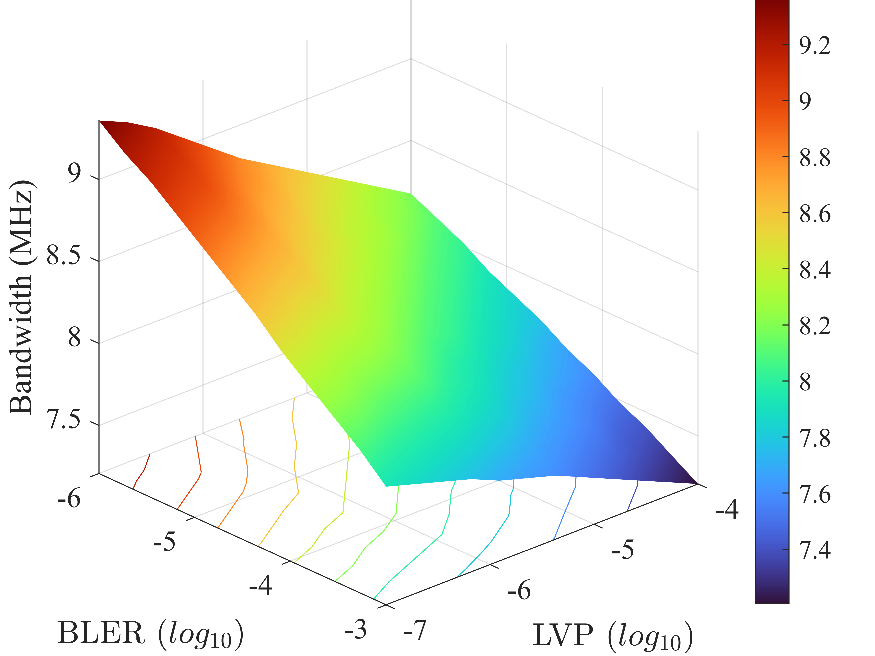}}
			\subfigure[optimal BER threshold]{
				\label{Fig:modelVa_RhoLVPBER_vTheta_2}
				\includegraphics[width=0.8\linewidth]{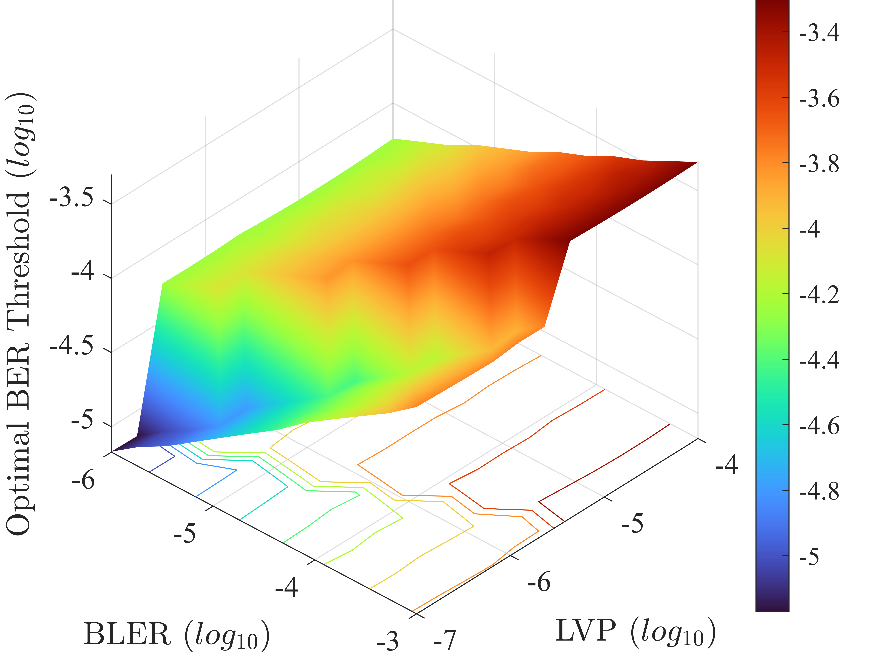}}
			\caption{Bandwidth consumption, the optimal BER threshold selection varying with the LVP and the BLER requirements, where the source data rate $\lambda=20$ Mbps, the total latency budget $D^{\text{th}}=10$ ms and the channel average SNR $\bar{\gamma} = 18$ dB.}
			\label{Fig:modelVa4}
		\end{figure}
		
		The joint effects of decoding BLER requirements and LVP constraints on optimal bandwidth consumption and optimal parameter selection are elucidated in Fig. \ref{Fig:modelVa_BLVPBER_vTheta_1} and Fig. \ref{Fig:modelVa_RhoLVPBER_vTheta_2}, respectively. It is evident that the bandwidth consumption under the same channel conditions and source rate requests decreases monotonically with decoding BLER and LVP. However, there are differences in the impact of decoding BLER requirements and LVP constraints on optimal parameter selection. The abrupt change in the optimal BER threshold along the LVP axis in Fig. \ref{Fig:modelVa_RhoLVPBER_vTheta_2} is attributed to the adjustment decision of the optimal transmission from two to three. Unlike the continuous variation of transmission parameters with decoding BLER requirements, the optimal BER threshold and transmission times selection do not change with LVP until the LVP requirements are relaxed enough to allow for a partial transfer of queuing time to transmission time, resulting in reduced bandwidth consumption, consistent with Fig. \ref{Fig:modelVa_RhoLVPGamma_vTheta_2}. Furthermore, for a lower BLER constraint, the increase in the maximum number of transmissions (i.e., the conversion from queuing time to transmission time) will occur at a smaller LVP. The above analysis verifies the rationality of the model and the integrated impact of the coupled parameters on system performance.
		
		\subsection{Performance Evaluation}
		\begin{figure}[tb]
			\centering
			\subfigtopskip=2pt 
			\subfigbottomskip=2pt 
			\subfigcapskip=-5pt 
			
			\subfigure[eMBB traffic]{
				\label{Fig:perfoEval1_eMBBTraffic}
				\includegraphics[width=0.8\linewidth]{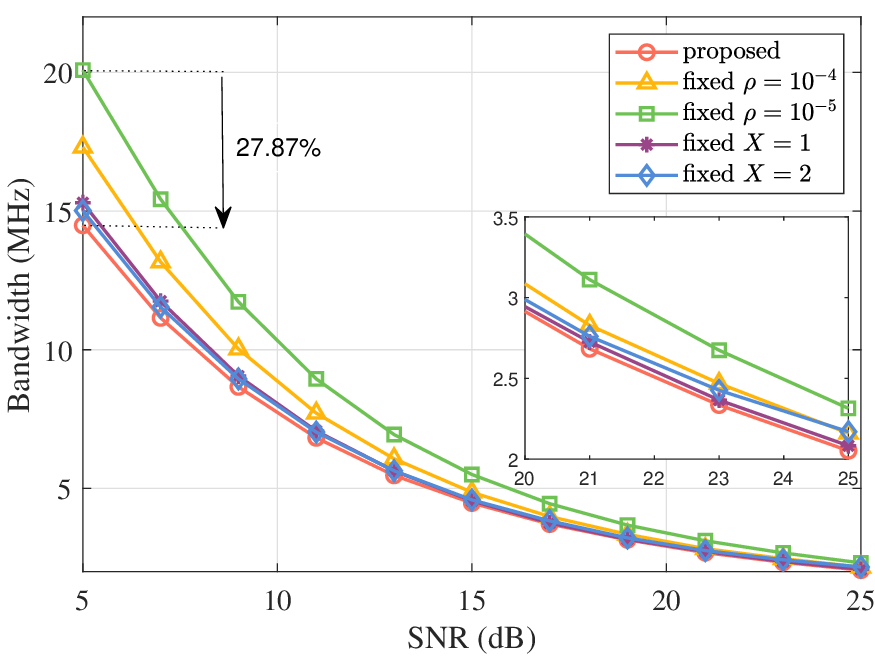}}
			\subfigure[uRLLC traffic]{
				\label{Fig:perfoEval1_uRLLCTraffic}
				\includegraphics[width=0.8\linewidth]{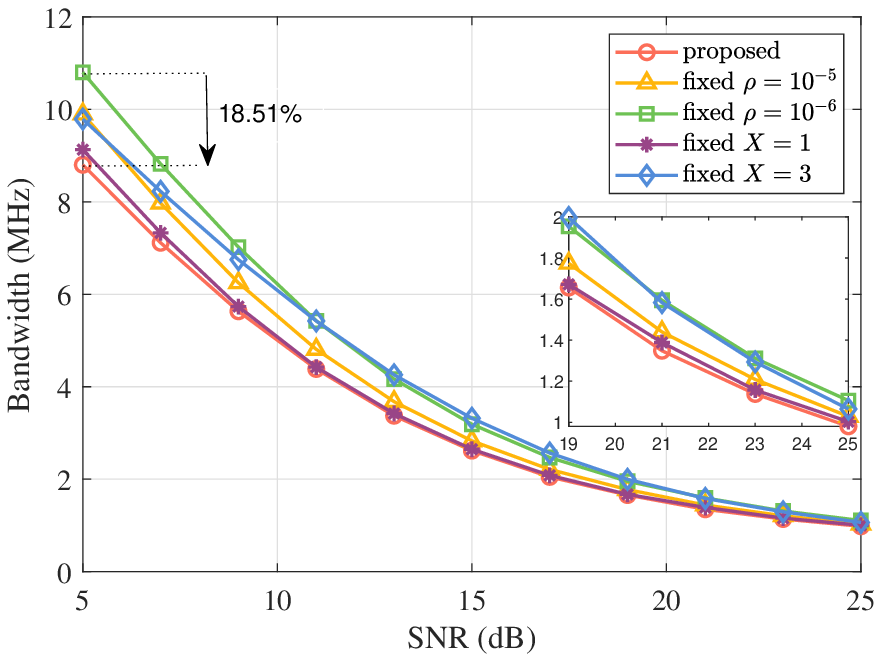}}
			\caption{Bandwidth consumption for ablation experiment under different channel qualities for eMBB and uRLLC traffic scenario, respectively.}
			\label{Fig:perfoEval1}
		\end{figure}
		
		In this section, we conduct a series of ablation experiments to evaluate the performance of the proposed cross-layer optimization framework. Specifically, transmission parameter is assigned fixed value in each layer separately, which is used in the current communication system. Joint optimization of parameters is compared with ablation schemes to demonstrate the optimality of our proposed algorithms. eMBB and uRLLC traffic are tested separately to evaluate the effectiveness of the algorithms for different types of traffic. Different delay budgets and LVP thresholds are set for two types of traffic based on their characteristics. For AP selection, $M=20$ APs and $N=40$ users are randomly distributed in a certain area, resulting in different channel qualities for each user when associated with different APs. A comparison between the auction algorithm based algorithm and the optimal channel quality based algorithm for AP selection is implemented to verify the balance between load balancing and bandwidth efficiency of the proposed algorithm.
		
		\begin{figure}[tb]
			\centering
			\subfigtopskip=2pt 
			\subfigbottomskip=2pt 
			\subfigcapskip=-5pt 
			
			\subfigure[Bandwidth vs. source rate of eMBB]{
				\label{Fig:perfoEval2_eMBBTraffic}
				\includegraphics[width=0.78\linewidth]{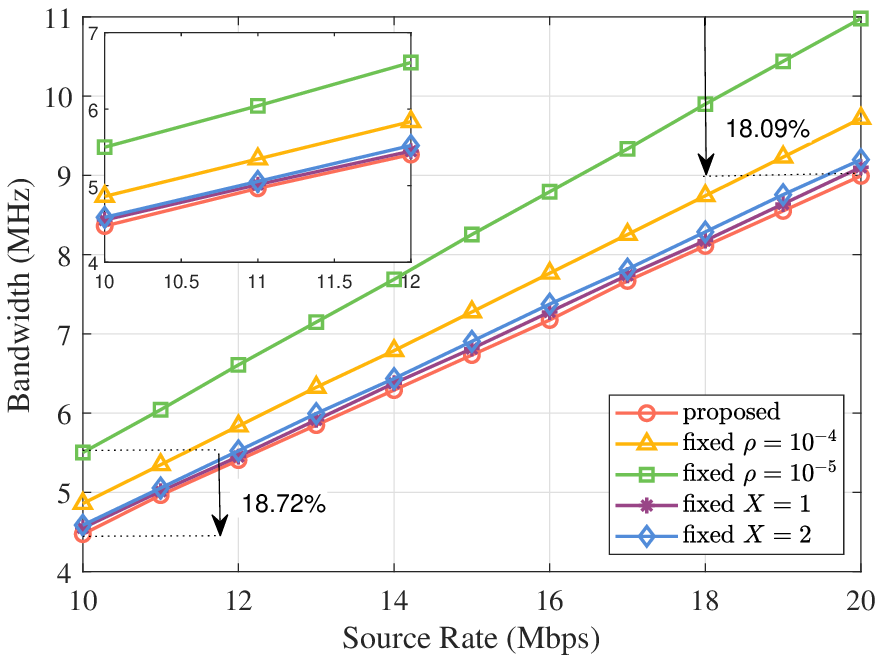}}
			\subfigure[Bandwidth vs. LVP of uRLLC]{
				\label{Fig:perfoEval3_uRLLCTraffic}
				\includegraphics[width=0.81\linewidth]{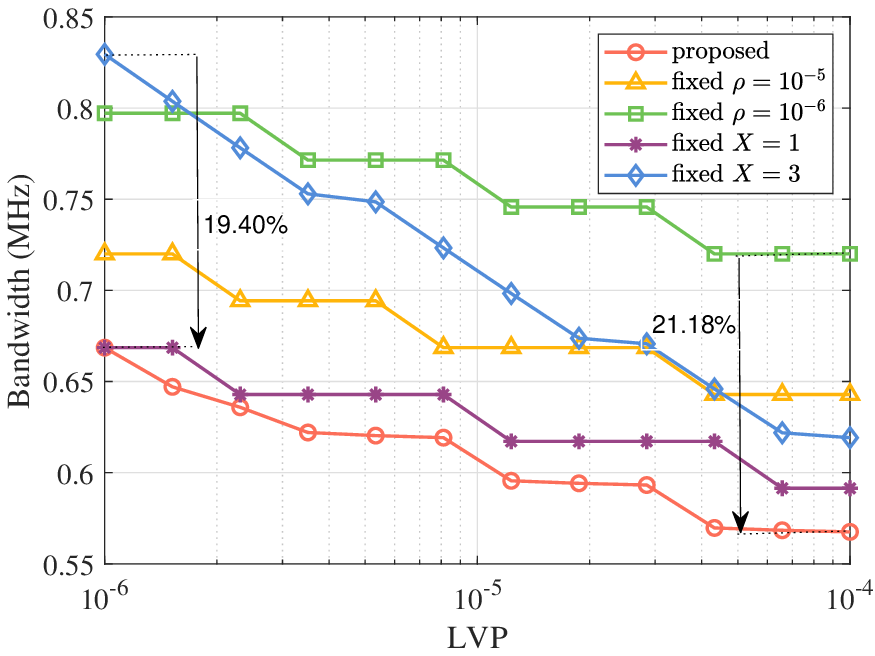}}
			\caption{Bandwidth consumption for ablation experiment under different source rate and LVP requirements for eMBB and uRLLC traffic scenario, respectively.}
			\label{Fig:perfoEval23}
		\end{figure}
		
		The bandwidth consumption when transmitting eMBB or uRLLC traffic through above TPD algorithm under different average channel qualities is shown in Fig. \ref{Fig:perfoEval1}. The rate requirement and latency budget for eMBB traffic are $\lambda=10$ Mbps and $D^{\text{th}}=10$ ms with LVP=$10^{-3}$, while for uRLLC traffic $\lambda=4$ Mbps and $D^{\text{th}}=5$ ms with LVP=$10^{-6}$. The bandwidth consumption caused by the proposed cross-layer scheme in different traffic scenarios and channel qualities is lower than when the transmission parameter is set to fixed value, which effectively verifies the superior performance of the algorithms. It can be obtained from Fig. \ref{Fig:perfoEval1_eMBBTraffic} that the bandwidth efficiency for eMBB traffic under proposed algorithms is up to $27.87\%$ higher than that of schemes with fixed transmission parameter, and up to $18.51\%$ higher in uRLLC scenario from Fig. \ref{Fig:perfoEval1_uRLLCTraffic}. Moreover, the intersection of bandwidth curves occurs in several fixed value schemes, which also proves the non optimality of fixed value schemes from the opposite side.
		
		The performance comparison results of the algorithms under different source rates and LVP requirements are illustrated in Fig. \ref{Fig:perfoEval23}, where source rate changes to simulate the eMBB traffic scenario and LVP changes to simulate the uRLLC traffic scenario. The average quality of the channel is set to the same $\bar{\gamma}=15$ dB in both scenarios. The parameters remain the same as before in the eMBB scenario, while for uRLLC traffic, the source rate requirement is reset to $\lambda=1$ Mbps. It is obvious that the proposed cross-layer scheme outperforms all other schemes equipped with fixed transmission parameters. Additionally, from Fig. \ref{Fig:perfoEval2_eMBBTraffic}, it can be observed that the higher delay budget of eMBB traffic leads to a higher optimal transmission count, which can achieve higher spectral efficiency by configuring a higher BER threshold. For uRLLC traffic, the optimal transmission count is 2 due to the lower delay budget, which is consistent with the analysis in the Sec. \ref{Sec: Model Validation}. The proposed algorithm can provide up to 18.72\% bandwidth efficiency gain in the EMBB scenario and up to 21.18\% in the uRLLC scenario compared with the schemes with fixed parameter. It is worth noting that the reason for partial curve in Fig. \ref{Fig:perfoEval3_uRLLCTraffic} being parallel to the LVP axis is the low bandwidth consumption of uRLLC traffic and the resource blocks-based bandwidth allocation.
		
		\begin{figure}[tb]
			\centering
			\subfigtopskip=2pt 
			\subfigbottomskip=2pt 
			\subfigcapskip=-5pt 
			\includegraphics[width=0.85\linewidth]{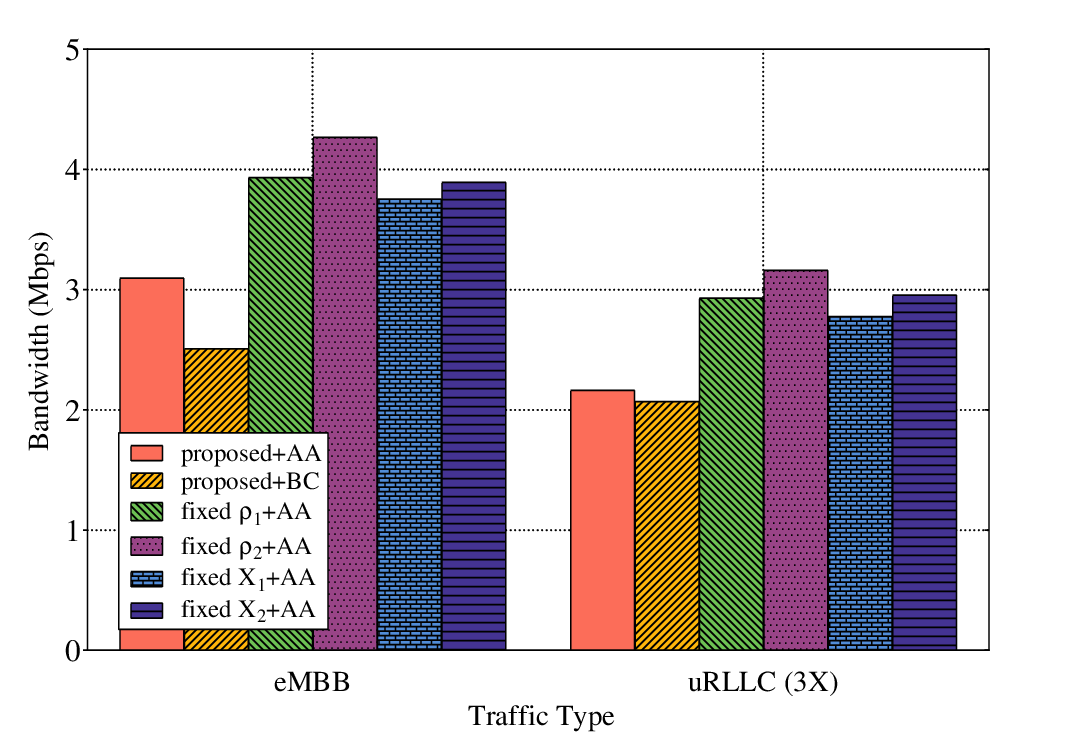}
			\caption{Comparison of joint performance of TPD algorithms and AP selection algorithms in different traffic scenarios.}
			\label{Fig:perfoEval4}
		\end{figure}
		
		By randomly placing $N=40$ users served by $M=20$ APs in a certain space, we simulate a realistic C-RAN scenario where half of the users request eMBB services and the other half request uRLLC services. For eMBB users, the source rate requirements and delay budgets are set to $[10,20]$ Mbps and $[10,16]$ ms, respectively, with LVP thresholds of $[10^{-4}, 10^{-3}]$. We set the source rate requirements to $[1, 5]$ Mbps for uRLLC traffic, with delay budgets $[3, 9]$ ms and LVP thresholds $[10^{-6}, 10^{-5}]$. The fixed parameters in the comparison scheme are consistent with the previous ablation experiments. Fig. \ref{Fig:perfoEval4} presents the performance comparison of different parameter configuration algorithms combined with AP decision algorithms under various service scenarios. It is worth noting that we enlarge the bandwidth in the uRLLC scenario to three times the original value for visual display. It can be seen that our proposed cross-layer parameter configuration algorithm achieves the best bandwidth efficiency performance in the auction-based algorithms (AA) for UA. On the other hand, the auction-based UA algorithm achieves the performance inferior to the best channel-based (BC) UA algorithm due to the higher spectral efficiency achievable with better channels, which is achieved at the expense of unbalanced load between APs. Therefore, our algorithm achieves a better balance between bandwidth efficiency performance and AP load balancing.

		\section{Conclusion}\label{Conclusion}
		In this paper, we have proposed a cross-layer parameters optimization and AP selection problem in C-RAN network to minimize the bandwidth consumption of system while satisfying statistical QoS requirements. We have designed the optimal cross-layer TPD algorithm for each possible AP-user pair, and then optimized the UA pattern through auction-based decision-making. The developed polynomial-time complexity algorithm can find the globally optimal transmission parameter configuration and AP selection, as the AP-user combinations are optimally selected from all possible pairs based on their optimal transmission parameters. Finally, numerical results validate the accuracy of the proposed QoS provision model and the effectiveness of the developed algorithm in terms of bandwidth savings.
		
		In the future, some meaningful researches including short-packet transmission and uRLLC in finite-length coding systems, are desirable of consideration. Analysis based on finite-length encoding is meaningful and important as we have mentioned, which will be more complicated in the C-RAN systems by jointly considering with packet type and QoS requirements. And it is worthy to find a more accurate representation of the spectral efficiency for more instructive performance results.
		
		\section*{Appendix A \\Proof of Theorem \ref{thm1}}
		The MCS selection scheme we adopt for the system is defined in \cite[Table 5.1.3.1-2]{3gpp_technical_38214}, which implies that the available spectrum efficiency $\upsilon_{j}$ is monotonically increasing with the MCS index $j$, that is, $\upsilon_{j+1} > \upsilon_{j}$ with a definite $\rho$. In addition, there is $\upsilon_{j}>0$ and $5\cdot\rho\ll 1$ due to definition and realistic characteristics respectively. So we have
		\begin{equation}
			\gamma_{j+1} > \gamma_{j}, \forall j \in \left\{0,1,\cdots,J\right\},
		\end{equation} 
		when $\rho$ is fixed and
		\begin{equation}\label{proof: gamma greater}
			\gamma_{j}\left( \rho^{\bullet} \right) < \gamma_{j}\left( \rho^{\circ} \right), \rho^{\bullet}>\rho^{\circ}, 0<\rho\ll 1.
		\end{equation}
		Similarly, we can conclude that $Pb\left(j,\gamma\right)$ and thus $P\left( 
		j,\gamma \right)$ decrease monotonically with $\gamma$ under same MCS $j$. We implement the following proofs based on the determined channel distribution with a fixed average SNR $\bar{\gamma}$.
		We study the function $F_{\gamma}\left(j,\gamma\right)$ with respect to the MCS $j$ and BER threshold $\rho$, which is defined as
		\begin{equation}\label{proof: Fgamajrho}
			\begin{aligned}
				F_{\gamma}\left(j,\rho\right) 
				&= \gamma_{j+1}\left(\rho\right) - \gamma_{j}\left(\rho\right) \\
				&= \frac{2}{3}\left(2^{\upsilon_{j}} - 2^{\upsilon_{j+1}}\right) \ln\left(5\cdot\rho\right).
			\end{aligned}
		\end{equation}
		Due to $\upsilon_{j+1} > \upsilon_{j}$, we can conclude that $F_{\gamma}\left(j,\gamma\right)$ decreases monotonically with $\rho$ when MCS fixes, which will be used later.
		
		First, we take a value $\rho^{\circ}$ and get the switching threshold $\gamma_{j}^{\circ}$ of MCS mode based on \eqref{switch threshold SNR}. And the BLER $\bar{P}(\rho^{\circ})$ can also be calculated by \eqref{actual BER}-\eqref{average Pcb}, given by
		{\setlength\abovedisplayskip{1pt}
			\setlength\belowdisplayskip{1pt}
			\begin{equation}\label{proof: Pcb0}
				\begin{aligned}
					\bar{P}(\rho^{\circ}) 
					&= \frac{1}{P_{T}^{\circ}} \sum_{j=0}^{J} \int_{\gamma_{j}^{\circ}}^{\gamma_{j+1}^{\circ}} P\left( j,\gamma \right) p_{\gamma}\left(\gamma\right) d\gamma \\
					&= \frac{1}{P_{T}^{\circ}} \sum_{j=0}^{J} \int_{\gamma_{j}^{\circ}}^{\gamma_{j+1}^{\circ}} \left[ 1\!\!-\!\!\left[1\!\!-\!\!Pb(j, \gamma)\right]^{L} \right] p_{\gamma}\left(\gamma\right) d\gamma,
				\end{aligned}
		\end{equation}} where $P_{T}^{\circ} = \int_{\gamma_{0}^{\circ}}^{+\infty} p_{\gamma}\left(\gamma\right) d\gamma$ is the probability that at least one MCS mode is available with BER threshold $\rho^{\circ}$. Then, we slightly increase the BER threshold to $\rho^{\bullet} = \rho^{\circ} + \epsilon$ with $0 < \epsilon \ll 1$ and consider corresponding BLER $\bar{P}(\rho^{\bullet)}$ with switching threshold $\gamma_{j}\left(\rho^{\bullet}\right)$, denoted by $\gamma_{j}^{\bullet}$ for simplicity, and probability $P_{T}^{\bullet}$. It's not difficult to obtain $\gamma_{j}^{\bullet}  < \gamma_{j}^{\circ}$ by \eqref{proof: gamma greater}, and thus we can conclude $P\left(j,\gamma_{j}^{\bullet}\right) > P\left(j,\gamma_{j}^{\circ}\right)$ for $\forall j \in \left\{ 0,1,\cdots,J \right\}$, where  $P\left(j,\gamma_{j}\right)$ denotes the BLER with MCS $j$ at switching threshold $\gamma_{j}$. In addition, we define the difference of MCS switching threshold between two BER thresholds under the same MCS as $\vartriangle\!\!\gamma_{j} = \gamma_{j}^{\circ} - \gamma_{j}^{\bullet}$, so we have
		\begin{equation}
			\begin{aligned}
				\vartriangle\!\!\gamma_{j+1} - \vartriangle\!\!\gamma_{j} 
				&= \left( \gamma_{j+1}^{\circ} - \gamma_{j+1}^{\bullet}\right) - \left(\gamma_{j}^{\circ} - \gamma_{j}^{\bullet}\right) \\
				&= \left[\gamma_{j+1}\left(\rho^{\circ}\right) - \gamma_{j}\left(\rho^{\circ}\right)\right] - \left[\gamma_{j+1}\left(\rho^{\bullet}\right) - \gamma_{j}\left(\rho^{\bullet}\right)\right] \\
				&\overset{\eqref{proof: Fgamajrho}}{=} F_{\gamma}\left(j,\rho^{\circ}\right) - F_{\gamma}\left(j,\rho^{\bullet}\right).
			\end{aligned}
		\end{equation}
		We can then obtain $\vartriangle\!\!\gamma_{j+1} - \vartriangle\!\!\gamma_{j}>0$ from the monotonicity of the $F_{\gamma}\left(j,\rho\right)$ with respect to $\rho$, which implies that the difference of MCS switching thresholds corresponding to the two BER thresholds increases monotonously with the MCS index $j$. 
		
		With MCS $j$, $Pb\left(j,\gamma\right)$ defined in $[\gamma_{j},\gamma_{j+1})$ takes the same value as $Pb\left(j,\gamma-\!\!\vartriangle\right)$ defined in $[\gamma_{j}+\!\!\vartriangle,\gamma_{j+1}+\!\!\vartriangle)$. Therefore, when MCS is fixed to $j$ and the $Pb^{\bullet}\left(j, \gamma\right)$ curve moves distance $\vartriangle\!\!\gamma_{j}$ in the positive direction of $\gamma$, we can get
		\begin{equation}\label{proof: conclude1}
			Pb^{\bullet}\left(j,\gamma-\vartriangle\!\!\gamma_{j}\right) > Pb^{\circ}\left(j,\gamma\right), \gamma \in [\gamma_{j}^{\circ}, \gamma_{j+1}^{\bullet}).
		\end{equation}
		In addition, we already know that $P$ decreases monotonically with respect to $\gamma$ with fixed MCS, so we have
		\begin{equation}\label{proof: conculde2}
			Pb^{\bullet}\left(j,\gamma_{j+1}^{\bullet}\right) > Pb^{\circ}\left(j,\gamma\right), \gamma \in [\gamma_{j+1}^{\bullet}, \gamma_{j+1}^{\circ}).
		\end{equation} 
		Based on \eqref{proof: conclude1} and \eqref{proof: conculde2}, we can conclude that
		\begin{equation}\label{roof: avPcb greater}
			\begin{aligned}
				&\frac{1}{P_{T,j}^{\bullet}} \int_{\gamma_{j}^{\bullet}}^{\gamma_{j+1}^{\bullet}} P\left( j,\gamma \right)p_{\gamma}\left(\gamma\right) d\gamma \\
				&\quad\quad > \frac{1}{P_{T,j}^{\circ}} \int_{\gamma_{j}^{\circ}}^{\gamma_{j+1}^{\circ}} P\left( j,\gamma \right)p_{\gamma}\left(\gamma\right) d\gamma,  \forall j \in \left\{ 0,1,\cdots, J \right\} ,
			\end{aligned}
		\end{equation} where $P_{T,j} = \int_{\gamma_{j}}^{\gamma_{j+1}} p_{\gamma}\left( \gamma \right) d\gamma$ denotes the probability of adopting MCS mode $j$. By summing the two sides of the \eqref{roof: avPcb greater} for $j\in\left\{0,1,\cdots,J\right\}$, we can obtain
		\begin{equation}
			\begin{aligned}
				&\sum_{j=0}^{J} \frac{1}{P_{T,j}^{\bullet}} \int_{\gamma_{j}^{\bullet}}^{\gamma_{j+1}^{\bullet}} \!\!P\left( j,\gamma \right)p_{\gamma}\left(\gamma\right) d\gamma \\
				&\quad\quad\quad\quad\quad\quad > \sum_{j=0}^{J} \frac{1}{P_{T,j}^{\circ}} \int_{\gamma_{j}^{\circ}}^{\gamma_{j+1}^{\circ}} \!\!P\left( j,\gamma \right)p_{\gamma}\left(\gamma\right) d\gamma.
			\end{aligned}
		\end{equation}
		The above formula can be rephrased as
		\begin{equation}
			\begin{aligned}
				&\frac{1}{P_{T}^{\bullet}} \sum_{j=0}^{J} \int_{\gamma_{j}^{\bullet}}^{\gamma_{j+1}^{\bullet}} P\left( j,\gamma \right) p_{\gamma}\left(\gamma\right) d\gamma \\
				&\quad\quad\quad\quad\quad\quad > \frac{1}{P_{T}^{\circ}} \sum_{j=0}^{J} \int_{\gamma_{j}^{\circ}}^{\gamma_{j+1}^{\circ}} P\left( j,\gamma \right) p_{\gamma}\left(\gamma\right) d\gamma,
			\end{aligned}
		\end{equation} which completes the proof by $\bar{P}\left(\rho^{\bullet}\right) > \bar{P}\left(\rho^{\circ}\right)$.

		\section*{Appendix B \\Proof of Theorem \ref{thm2}}
		To prove this theorem, it is sufficient to demonstrate the monotonicity of functions $\psi_{j}\left(\rho\right)$ and $\theta\left(\rho\right)$ with respect to $\rho$. First, the average information per RB $\mathbb{E}_{\gamma}\left[\psi\right]$ in the ergodic channel can be calculated as
		\begin{equation}
			\begin{aligned}
				\mathbb{E}_{\gamma}\left[\psi\right]
				&=\frac{1}{P_{T}} \sum_{j=0}^{J} \int_{\gamma_{j}}^{\gamma_{j+1}} \psi_{j}\cdot p_{\gamma}\left(\gamma\right) d\gamma \\
				&=\frac{1}{P_{T}} \sum_{j=0}^{J} \int_{\gamma_{j}}^{\gamma_{j+1}} \alpha\cdot\beta\cdot\upsilon_{j}\cdot p_{\gamma}\left(\gamma\right) d\gamma,
			\end{aligned}
		\end{equation} where the definition of $P_{T}$ is consistent with the body. Similar to \eqref{proof: conclude1} and \eqref{proof: conculde2} in the proof of property \ref{thm1}, we have $\upsilon_{j+1}^{\bullet} > \upsilon_{j}^{\circ}$ for $\gamma \in [\gamma_{j}^{\bullet}, \gamma_{j}^{\circ}]$ and $\upsilon_{j+1}^{\bullet} = \upsilon_{j+1}^{\circ}$ for $\gamma \in [\gamma_{j+1}^{\circ}, \gamma_{j+2}^{\bullet}]$. Therefore, $\mathbb{E}_{\gamma}\left[\psi\right]$ is increasing with BER threshold $\rho$ since average spectrum efficiency $\mathbb{E}_{\gamma}\left[\upsilon\right]$ increases with $\rho$. 
		
		From the expression of $\theta\left(\rho\right)$ in \eqref{theta calculation}, we can obtain that $\theta\left(\rho\right)$ decreases monotonically with $\rho$ if \eqref{optimal EC Constraint} holds. Further, since the effective capacity in \eqref{F_mn} is increasing with $\rho$ due to similar reason with $\mathbb{E}_{\gamma}\left[\upsilon\right]$, we can conclude that $F_{m,n}\left(\rho_{m,n}\right)$ is monotonically increasing with $\rho_{m,n}$, which completes the proof.

		\bibliographystyle{IEEEtran}
		\bibliography{202302}
		
	\end{document}